\documentclass[%
 amsmath,amssymb,
 aps,
]{revtex4-2}

\usepackage{graphicx}
\usepackage{dcolumn}
\usepackage{bm}
\usepackage{hyperref}

\begin{document}

\title{Temperature Measurement on Copper Surfaces for Superconducting Thin Film Cavity Applications}

\author{Antonio Bianchi} \email{Corresponding author: antonio.bianchi@cern.ch}
\author{Giovanna Vandoni}
\author{Walter Venturini Delsolaro}%
\affiliation{CERN}

\begin{abstract}

Superconducting radio-frequency (SRF) thin film cavities on copper substrates are employed in several particle accelerators. However, these SRF cavities historically featured a progressive performance degradation with the accelerating field that is still not completely understood. The degradation of cavity performance, which limits the use of this technology in accelerators where the real-estate gradient has to be maximized, is manifested by the presence of heat losses in the superconducting film. However, measuring the temperature on the outer surface of copper substrates is challenging due to the higher thermal conductivity of copper at low temperatures compared to niobium.

This study describes how temperature variations on copper surfaces can be satisfactorily measured in view of superconducting thin film cavity applications at liquid helium temperatures. Furthermore, we explore how the thermal exchange between thermometers and copper surfaces, and thermometers and helium bath must be tuned with respect to each other in order to measure accurately temperature rises in the thin film. Our findings suggest that engineering the copper surfaces can improve heat transfer into the helium bath and potentially enhance the performance of thin film SRF cavities.

\end{abstract}

\maketitle


\section{\label{sec:level1}INTRODUCTION}

For many applications to particle accelerators, superconducting thin film radio-frequency (RF) cavities are an interesting alternative to bulk niobium (Nb) ones, which are nowadays cutting-edge technology objects made from high purity Nb sheets welded together by electron-beam welding with high precision \cite{padamsee2}. Superconducting thin film RF cavities are usually built from a copper (Cu) substrate that is internally coated by a 1$-$2 $\mu$m thin film in Nb \cite{benvenuti1984niobium}. Not only do thin film RF cavities deposited on Cu substrates allow reducing the costs of raw materials, but they also have several advantages in comparison to bulk Nb cavities. Indeed, the substrates in Cu provide excellent stability to prevent thermal runaway of cavities and their usage can mitigate microphonics \cite{padamsee2, padamsee}. Moreover, the coating parameters of the superconducting thin film can be tuned to minimize its BCS surface resistance \cite{halbritter1974surface}. In addition, thin films often have a low sensitivity to the trapped magnetic field \cite{aull, miyazaki2019two}, which makes the design of cryomodules easier and cheaper because no magnetic shield is generally required \cite{Bruning:782076}.

Thin film coated Cu cavities are applied in several particle accelerators. Since the late '80s, CERN has pioneered the development of Nb/Cu cavities that have been successfully used in LEPII \cite{benvenuti1991superconducting, chiaveri1995production, schirm1995analysis}, LHC \cite{chiaveri1999cern} and HIE-Isolde linac \cite{venturini2013nb}. However, thin film cavities deposited on Cu historically featured a substantial increase in surface resistance with the accelerating field \cite{benvenuti1999study, miyazaki2019two}, resulting in a quality factor decrease (Q-slope). This is in part still unexplained. Unfortunately, the degradation of the quality factor in thin film cavities does not allow using this technology in accelerators where the real-estate gradient has to be maximized \cite{longmarch}.

Performance degradation in bulk Nb cavities is generally correlated to heat losses in their inner surface \cite{padamsee}. One of the most useful diagnostic tools to investigate the mechanisms responsible for performance degradation in superconducting radio-frequency (SRF) cavities is a temperature mapping system \cite{padamsee2020history, ciovati2005temperature, piel1981cern, bernard1981experiments, bernard1980first, pekeler1996thermometric, makita2015temperature}. This system, extensively applied in testing bulk Nb cavities, allows measuring the temperature on the outer surface of cavities and, consequently, permits the detection of internal point-like and extended dissipation regions during cavity cold tests.

Temperature measurements in thin film coated Cu cavities are challenging, unlike in bulk Nb cavities. Indeed, the thermal conductivity of Cu is more than one order of magnitude higher than that of Nb at liquid He temperatures \cite{padamsee1983calculations, russenschuck2011field}. The thermal conductivity of Cu, generally used for SRF cavities, is $\sim$300 W/(m$\cdot$K) at 1.8 K and $\sim$700 W/(m$\cdot$K) at 4.2 K, whereas these values in bulk Nb surfaces are only $\sim$25 W/(m$\cdot$K) and $\sim$60 W/(m$\cdot$K), respectively. This implies that the temperature profile on Cu surfaces in correspondence with a given heat loss is much lower than that on Nb surfaces.

In this study, we report how temperature variations on Cu surfaces can be satisfactorily measured in view of developing a temperature mapping system for thin film coated Cu cavities. We describe how the thermal exchange between thermometers and Cu surfaces, and thermometers and helium (He) bath, plays a crucial role in measuring temperature rises on Cu surfaces. In addition, we demonstrate that engineering conveniently the outer surface of Cu surfaces improves the heat dissipation into the He bath and, consequently, may imply increased performance in thin film coated Cu cavities. 

The paper is organized as follows. After the description of the experimental set-up and the characterization of thermometers in section \ref{sec:level2}, we examine in section \ref{sec:level3} how to reduce the thermal contact resistance between thermometers and Cu surfaces in order to improve the thermometer response in the presence of heat losses. In section \ref{sec:level4} we describe how the roughness of Cu surfaces modifies the temperature rise in the presence of a point-like heat loss. Section \ref{sec:level5} examines the temperature profile on Cu surfaces in the presence of heat losses at different conditions and temperatures of the He bath. In addition, we report in section \ref{sec:level8} the temperature profiles along a Nb/Cu 1.3 GHz elliptical cavity cell measured during a vertical cold test. Finally, results are discussed in section \ref{sec:level9}.

\section{\label{sec:level2}EXPERIMENTAL SET-UP}

\begin{figure}[!htb]
   \centering
   \includegraphics*[width=0.35\columnwidth]{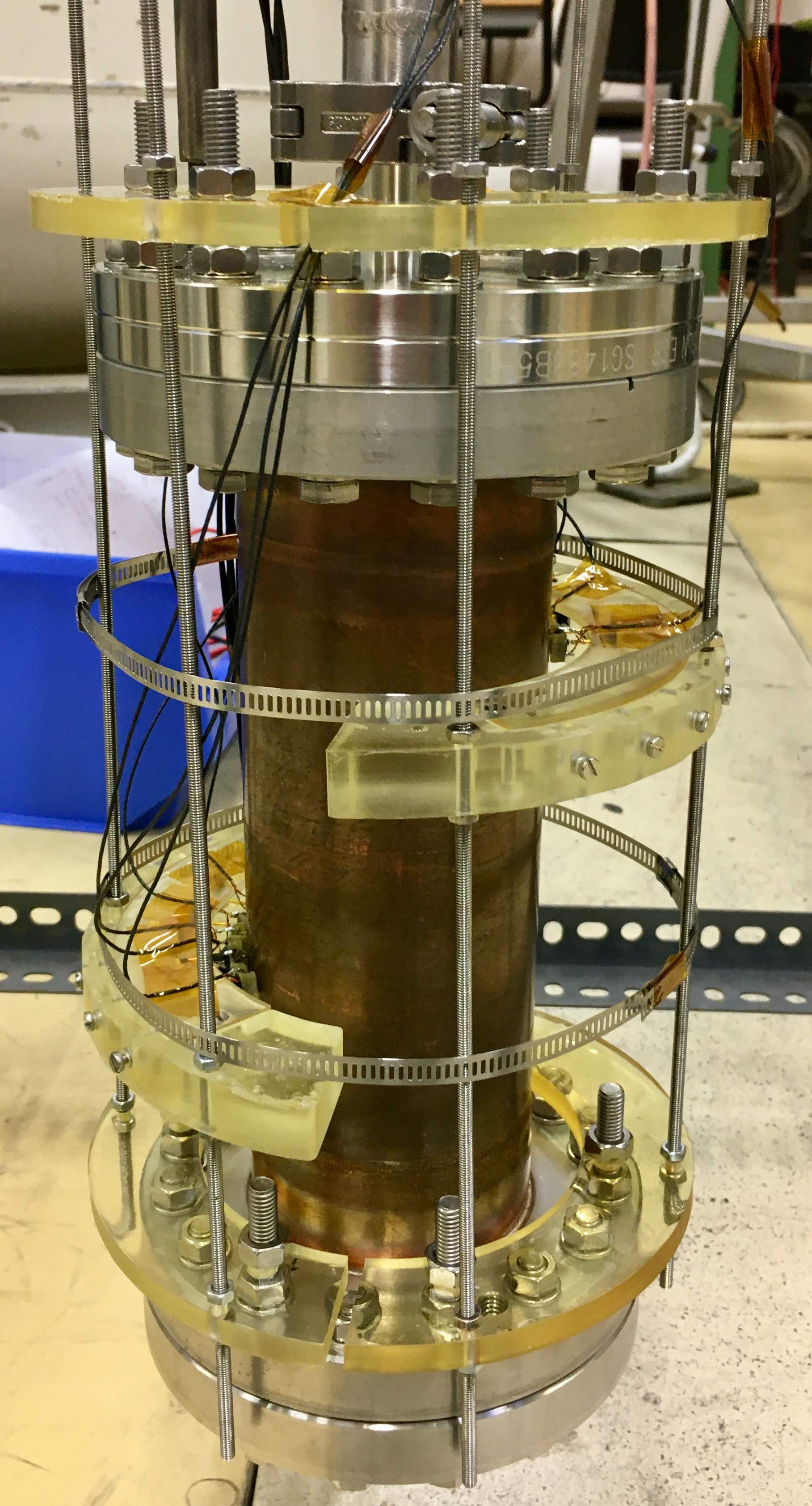} 
   \caption{Photo of the experimental set-up: tube in OFE Cu with a RRR of $\sim$50 surrounded by a supporting system in Araldite MY750 for pushing the thermometers towards the outer surface of the Cu tube.}
   \label{fig:photo_Cutube}
\end{figure}

\begin{figure}[!htb]
   \centering
   \includegraphics*[width=0.35\columnwidth]{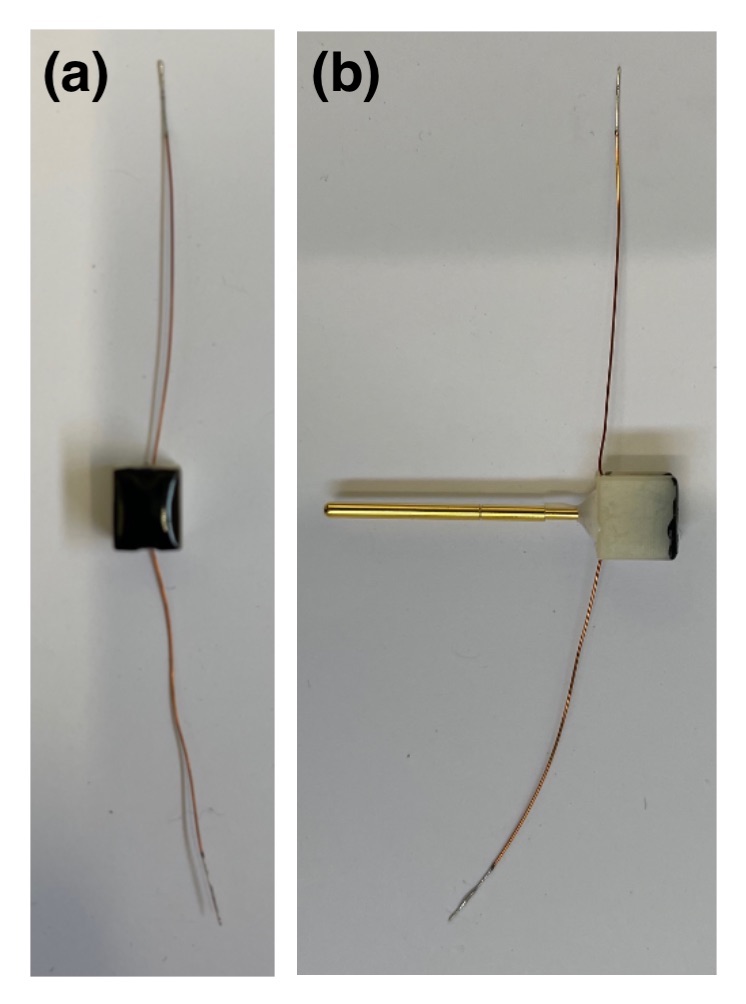} 
   \caption{Top view (a) and side view (b) of a thermometer embedded in an Accura 25 housing (in white) and sealed with Stycast epoxy (in black). A spring-loaded pin (yellow) is glued in the Accura 25 housing by silicone rubber CAF4. Two wires in manganin (brown) are welded to the thermometers.}
   \label{fig:photo_thermometerAccura}
\end{figure}

In order to carry out temperature measurements on Cu surfaces, we used an 8 cm diameter tube in OFE Cu with a wall thickness of 2 mm and a residual resistivity ratio (RRR) of $\sim$50. This type of Cu is widely used in superconducting thin film cavity applications. Two different heaters are glued in the inner surface of the Cu tube with Stycast 2850FT epoxy, which has a high thermal conductivity at liquid He temperatures. One heater is a thick film surface-mount (SMD) resistor of 30 $\Omega$ at 300 K, whereas the other is a rectangular and flexible heater in polyimide with a total resistance of 8 $\Omega$ at 300 K. By using these two heaters, we can reproduce point-like heat losses as well as extended dissipation regions. Thermometers are placed outside the tube and pushed towards the outer surface of the Cu tube using a supporting system in Araldite MY750, as shown in figure \ref{fig:photo_Cutube}, and spring-loaded pins (pogo-sticks) in BeCu, shown in figure \ref{fig:photo_thermometerAccura}b.

Thermometers are embedded in an Accura 25 housing and sealed with Stycast epoxy, which is impervious to superfluid He \cite{padamsee2020history}, as shown in figures \ref{fig:photo_thermometerAccura}a and \ref{fig:photo_thermometerAccura}b. Our tests have shown that Accura 25, a 3D-printed plastic material, is suitable for use at low temperatures. Wires from each thermometer are in manganin that is characterized by a low thermal conductivity at low temperatures, whereas the spring-loaded pins are glued in the Accura 25 housing by silicone rubber CAF4.

Voltage signals of thermometers are acquired by the 4-wire sensing method and digitized by the National Instruments 9251 ADCs with a sampling rate of 1 kS/s. Two Keithley 2401 modules provide the power supply. With a National Instruments 9472 module and a few lead batteries of 6 V, voltage pulses are generated to fire heaters at different power values and for various time durations.

Different types of thermometers and thermal contacts were tested with the Cu tube completely immersed in a liquid He bath. The internal pressure of the tube is constantly kept lower than $\sim$10$^{-4}$ Pa during measurements. The Cu surfaces are always vertically oriented during tests in this study. 

\begin{figure}[!htb]
   \centering
   \includegraphics*[width=0.65\columnwidth]{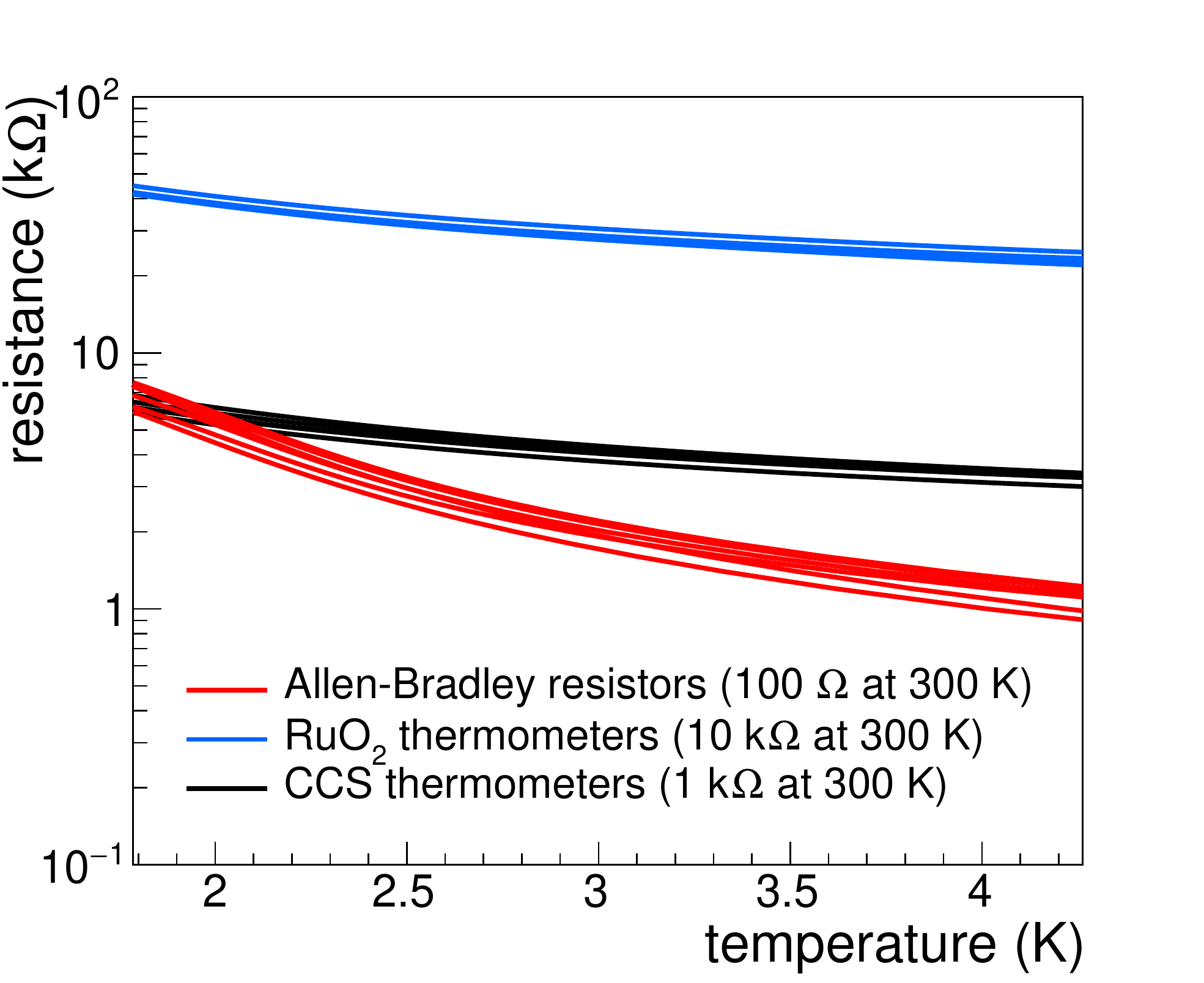} 
   \caption{Resistance variation of Allen-Bradley resistors, RuO$_{2}$ thermometers and carbon ceramic sensors (CCS) between 1.9 K and 4.2 K.}
   \label{fig:plot0}
\end{figure}

The experimental setup described in this work allows us to characterize different types of thermometers and evaluate their resistance variation as a function of temperature. We tested Allen-Bradley resistors with a resistance of $\sim$100 $\Omega$ at 300 K, ruthenium oxide (RuO$_{2}$) thermometers of $\sim$10 k$\Omega$ at 300 K, and Carbon Ceramic Sensors (CCS) of $\sim$1 k$\Omega$ at 300 K. The bias current for Allen-Bradley resistors is set at 10 $\mu$A as previous studies have shown that their self-heating is negligible for currents lower than 25 $\mu$A \cite{conway2017instrumentation, canabal2007development}. Likewise, the current for the other two types of thermometers is kept equal to 10 $\mu$A. The temperature of the He bath is measured by one calibrated Temati carbon ceramic sensor, immersed in the bath, and used as a reference. Figure \ref{fig:plot0} shows the variation of resistance in Allen-Bradley resistors, RuO$_{2}$ thermometers, and carbon ceramic sensors between 1.9 K and 4.2 K.

The resistance of Allen-Bradley resistors at 1.9 K is 6$-$8 times higher than that at 4.2 K. On the contrary, this resistance variation is only $\sim$1.5 times for RuO$_{2}$ thermometers and carbon ceramic sensors. 
The Allen-Bradley 100 $\Omega$ resistors were selected for this study due to their highest sensitivity in the temperature range of our interest, making them suited for measuring temperatures on Cu surfaces. Thermometers ensure a satisfactory resolution in measuring temperature profiles, as shown in the following sections. As reported in the literature \cite{ciovati2005temperature}, the resistance of Allen-Bradley resistors slightly varies each time they are cooled from room temperature to liquid He temperatures. We estimated that the variation of resistance from one thermal cycle to the next could induce uncertainties in temperature measurements higher than 150 mK. Therefore, all thermometers must be calibrated each time the experimental set-up is cooled down to liquid He temperatures before taking data. All temperature measurements shown in this study are taken by Allen-Bradley 100 $\Omega$ resistors previously calibrated between 1.9 K and 5.0 K.

With this experimental setup, we can evaluate how the thermal exchange between thermometers and Cu surfaces, and thermometers and He bath, have to be tuned with respect to each other in order to carry out temperature measurements on Cu surfaces. Indeed, not only does the sensitivity of thermometers change between 1.9 K and 4.2 K, but also the thermal conductivity of Cu and He varies in this temperature range. For this reason, temperature profiles on Cu surfaces are measured in this study at 1.9 K and 4.2 K at saturated vapor pressure, where the He is superfluid (He-II) and normal liquid (He-I), respectively, but also at 2.4 K at saturated vapor pressure and in subcooled He. Indeed, at 2.4 K, the thermal conductivity of Cu with a RRR of $\sim$50 is almost half of that at 4.2 K while the thermal conductivity of He decreases from $\sim$2.7 W/(m$\cdot$K) at 4.2 K to $\sim$1.85 W/(m$\cdot$K) at 2.4 K. Measurements at 2.4 K are taken at both saturation pressure and in subcooled condition, because the heat transfer between Cu and He considerably changes, as explained in details in section \ref{sec:level3}.

Studies on thermal contact resistance between thermometers and Cu surfaces are described in section \ref{sec:level3}, whereas the roughness effect on temperature measurements is presented in section \ref{sec:level4}. Finally, the heat transfer from Cu surfaces to He bath is examined in section \ref{sec:level5}.

\section{\label{sec:level3}THERMAL CONTACT}
The thermal contact resistance between thermometers and the Cu surface has been optimized to increase the resolution of temperature measurements as much as possible. Indeed, the thermal contact resistance affects the heat conduction from the outer surface of the Cu tube to the thermometers; it depends on several parameters, in particular, the roughness of the interface and the force holding the two surfaces together \cite{salerno1997thermal, salerno1984thermal, salerno1994thermal}. Optimizing thermal contact resistance is fundamental to detecting temperature variations on Cu surfaces with a satisfactory resolution.

Different solutions were tested to decrease the thermal contact resistance of the interface. The contact force has been maintained approximately constant using spring-loaded pins that push the thermometers toward the Cu surface. This ensures that thermometers remain in contact with the Cu surface during the cooling of the system from room temperature to liquid He temperatures. Indeed, the spring-loaded pins compensate for the distance gap between thermometers and the Cu surface caused by the thermal contraction of materials during the cooling.

If thermometers are in contact with the Cu surface without any additional thermal paste in the interface, the thermal contact resistance is high enough that no temperature variation is detected, except for subcooled He at $\sim$2.4 K. Therefore, double-sided adhesive tape in Kapton polyimide, indium foil, Apiezon N grease and silicone rubber CAF4 charged with copper powder have been tested in the interface between thermometers and the Cu surface with the aim to improve the thermal contact. All these materials are suitable for cryogenic temperatures. However, satisfactory results are only obtained with Apiezon N grease. 

\begin{figure*}[!htb]
   \centering
   \includegraphics*[width=0.95\textwidth]{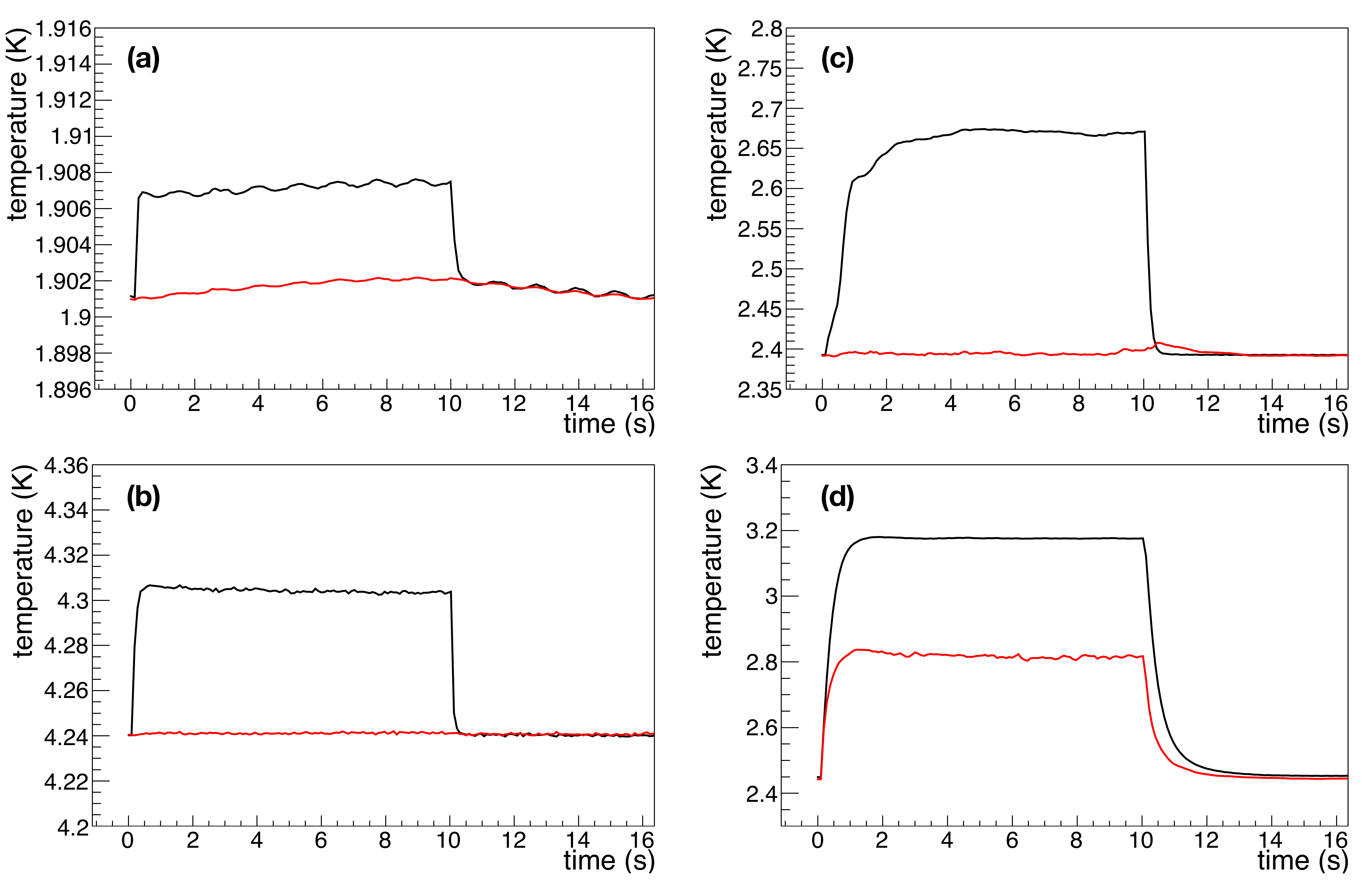}
   \caption{Temperature variation detected by two thermometers when a 9.1 x 3.8 cm$^{2}$ heater is fired at 80 mW/cm$^{2}$ for 10 s. Measurements are taken at 1.9 K (a), 4.2 K (b), 2.4 K (c) at saturation pressure and $\sim$2.4 K in subcooled He (d). The response of the thermometer in contact with the Cu surface by Apiezon N grease is in black, whereas the response of the thermometer without thermal paste in the interface is in red.}
   \label{fig:plot1e2e3e4}
\end{figure*}

Measurements have been carried out by placing two thermometers in contact with the outer surface of the Cu tube in the middle of a wide area that is heated on the opposite side of the surface by the rectangular heater. The heater has a total area of 9.1 x 3.8 cm$^{2}$. The thermometers, spaced 1 cm from each other, are in correspondence with the center of the rectangular heater where the heat flux is uniform. The spring-loaded pins press both thermometers against the surface of the Cu with a similar amount of force, ensuring comparable contact force. One thermometer is in contact with the Cu surface without anything in the interface, whereas a layer of Apiezon N grease is added to the second thermometer. Figures \ref{fig:plot1e2e3e4} show the temperature variations detected by both thermometers when the heater is fired at 80 mW/cm$^{2}$ for 10 s. The temperature of the He bath in figures \ref{fig:plot1e2e3e4}a, \ref{fig:plot1e2e3e4}b and \ref{fig:plot1e2e3e4}c is 1.9 K, 4.2 K, and 2.4 K at saturation pressure, respectively.

If Apiezon N grease is not present in the interface, no temperature increases are detected by thermometers at 1.9 K, 2.4 K, and 4.2 K at saturation pressure, even when the heater is fired for more than 30 s up to a heat flux density of 1.5 W/cm$^{2}$ that is the maximum value reachable with our experimental set-up. On the contrary, a temperature increase is detected in all three He bath conditions if a layer of Apiezon N grease is added to the interface. Indeed, the Apiezon N grease significantly reduces the thermal contact resistance and partially shields the thermometers from the liquid He \cite{padamsee2020history}, permitting the direct measurement of a fraction of the surface temperature at 1.9 K, 2.4 K, and 4.2 K. The efficiency of thermometers, defined as the ratio of the measured temperature rise to the theoretical temperature rise, is reported for different temperatures of the He bath in section \ref{sec:level7}.

In addition to the He bath conditions of 1.9 K, 2.4 K, and 4.2 K at saturation pressure, we also carried out some tests in subcooled He where the bath is at $\sim$2.4 K while the pressure in the cryostat is intentionally kept equal to the atmospheric pressure ($\sim1\cdot 10^{5}$ Pa). The overpressure in the cryostat partially inhibits the formation of He bubbles in the bath of liquid He. As a consequence, the cooling capability of the He bath is much lower than that at saturated vapor pressure, which is $\sim$80 mbar at 2.4 K. This generally increases the temperature of the Cu surfaces in the presence of heat losses, and the thermometer response turns out to be much higher in a subcooled He bath, as reported by H. Piel \cite{piel1980diagnostic, piel1981cern}. Figure \ref{fig:plot1e2e3e4}d shows the thermometer response in subcooled He (at $\sim$2.4 K with an overpressure of $\sim1\cdot 10^{5}$ Pa) with and without Apiezon N grease in the interface when the heater is fired at 80 mW/cm$^{2}$ for 10 s. Unlike the cases at 1.9 K, 2.4 K, and 4.2 K at saturation pressure where a temperature increase of the surface can be only detected using Apiezon N grease, the direct measurement of a fraction of the surface temperature in subcooled He is possible even without grease. Indeed, the signal of the thermometer without Apiezon N grease corresponds to a temperature increase of $\sim$350 mK, as shown in figure \ref{fig:plot1e2e3e4}d, but it is increased by more than 100\% with the usage of Apiezon N grease. As a consequence, the temperature of Cu surfaces can be satisfactorily measured in subcooled He without grease in the interface, but the thermometer response in this configuration is lower than that with the usage of the grease.

No significant difference was found by exchanging the thermometers and repeating the measurement campaign. This confirms that measurements are affected by neither the single thermometer nor by its position relative to the heater.

Hereafter, only results acquired by Allen-Bradley 100 $\Omega$ resistors with Apiezon N grease in the interface are reported unless otherwise specified.

\section{\label{sec:level4}ROUGHNESS}

The roughness of the surface plays an important role in the heat exchange from Cu surfaces to He bath, as described in \cite{smith1969review, bald1976nucleate, schmidt1981review}. This is especially observed when the Cu surface is in a nucleate boiling regime. A surface with higher roughness generally has a higher density of potential nucleation sites where the formation of He bubbles can occur. The activation of nucleation sites in the surface leads to an enhanced heat transfer rate compared to that of only convection cooling. According to \cite{bald1976nucleate, schmidt1981review}, a rougher surface seems to cause an increase in the heat transfer coefficient with a negligible effect on the peak nucleate boiling flux. Therefore, the roughness of the surface is a parameter that should be considered for superconducting thin film cavity applications because it can affect the temperature increase of cavities, especially in the presence of hotspots. 

In this study, the thick film SMD resistor, glued to the internal surface of the Cu tube, has been used as a heater to simulate a potential point-like heat loss in the Cu tube. The heater is chosen with a total area of $\sim$1 cm$^{2}$. Two thermometers are symmetrically placed to the left and right of the heater. One thermometer is glued to the internal surface of the Cu tube, while the second thermometer is pressed against the outer surface by a spring-loaded pin. The distance of both thermometers from the heater is 1 cm. With this configuration, measurements of temperature variations as a function of power can be carried out without affecting the response of each thermometer. If the external thermometer had been placed in correspondence with the internal one, the presence of the first thermometer would have affected the thermometer response of the second one and vice versa. To minimize this issue, the internal and external thermometers have been symmetrically placed to the left and right of the heater.

\begin{figure}[!htb]
   \centering
   \includegraphics*[width=0.8\columnwidth]{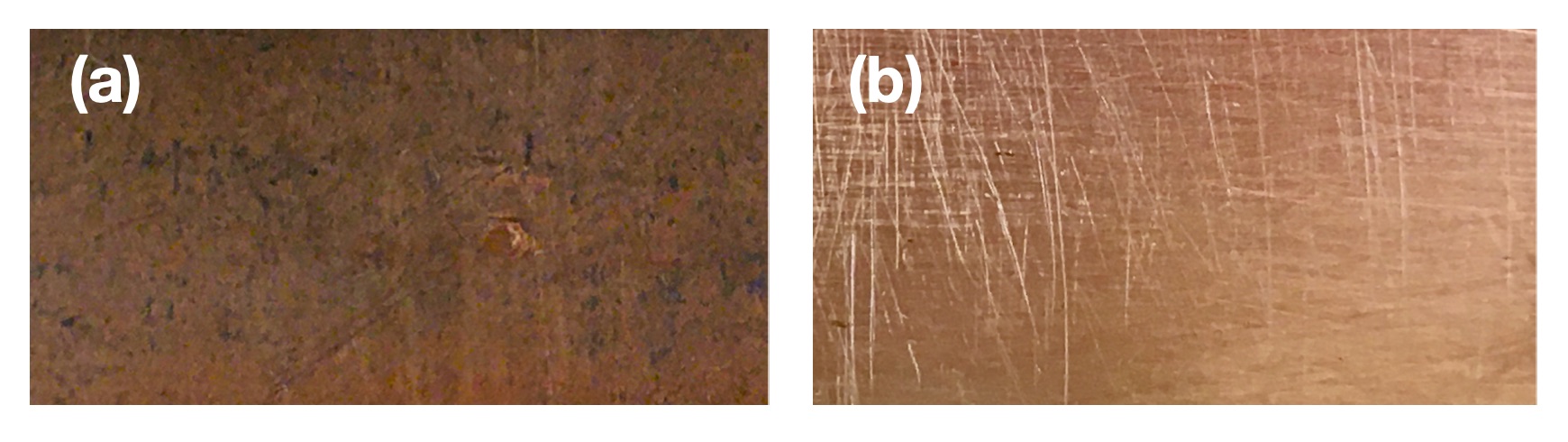}
   \caption{Smooth (a) and rough (b) surface in OFE Cu (RRR of $\sim$50).}
   \label{fig:photo_surfaceflatandrough}
\end{figure}

Temperature variations were measured on the Cu tube when its external surface was smooth. Then, the Cu surface was scratched with 80-grit sandpaper in order to increase its roughness, and measurements were repeated. Figure \ref{fig:photo_surfaceflatandrough}a shows the smooth Cu surface, while a photo of the rough surface is shown in figure \ref{fig:photo_surfaceflatandrough}b. The color of the Cu surface in figure \ref{fig:photo_surfaceflatandrough}a is darker than that in figure \ref{fig:photo_surfaceflatandrough}b as the smooth surface is more oxidized.

\begin{figure}[!htb]
   \centering
   \includegraphics*[width=0.65\columnwidth]{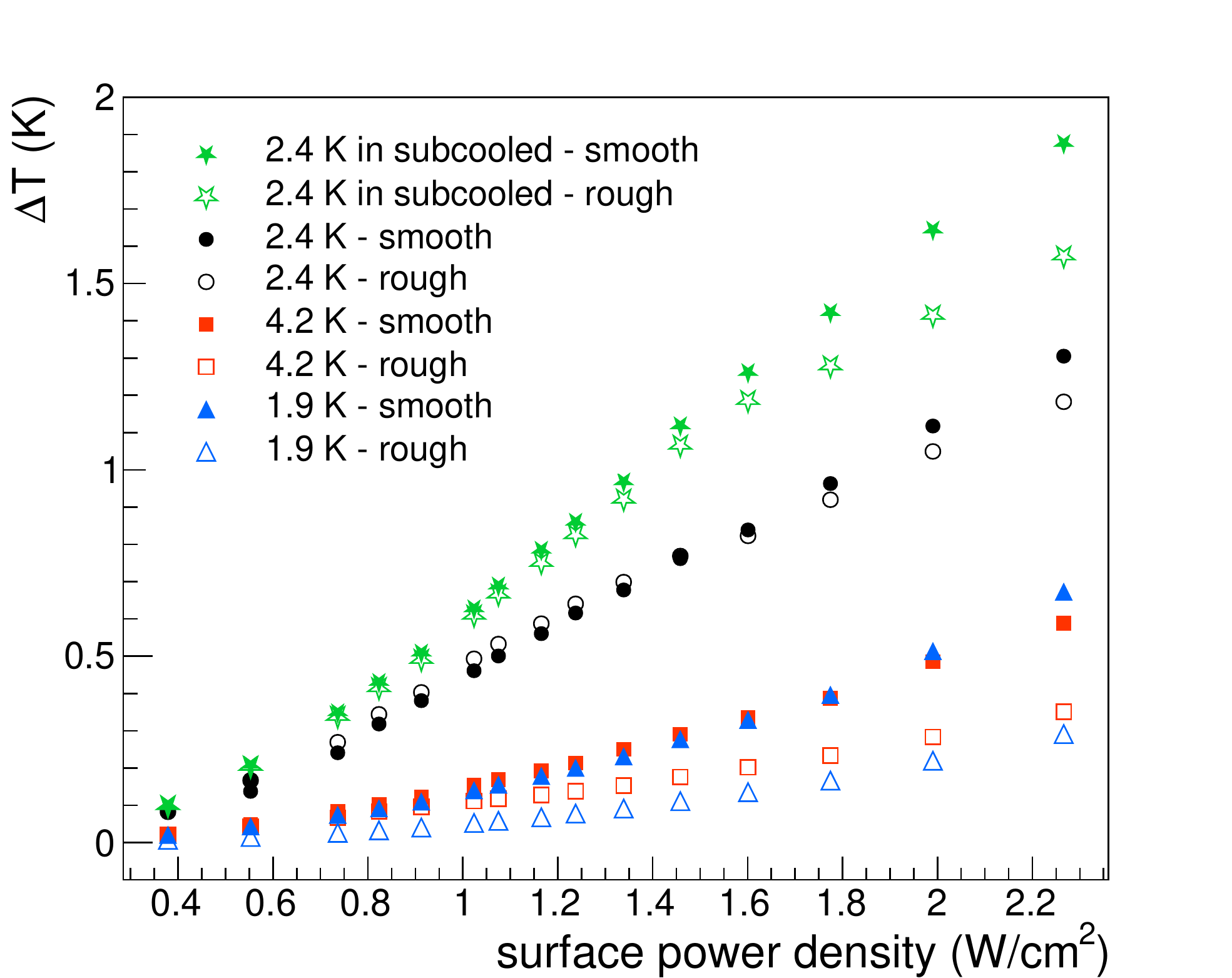} 
   \caption{Temperature variations measured by the internal thermometer at 1 cm from the heater on the smooth and rough Cu surfaces at 1.9 K (blue), 2.4 K (black), 4.2 K (red) at saturation pressure, and $\sim$2.4 K in subcooled He (green). Some statistical error bars are hidden by markers.}
   \label{fig:plot5}
\end{figure}

\begin{figure}[!htb]
   \centering
   \includegraphics*[width=0.65\columnwidth]{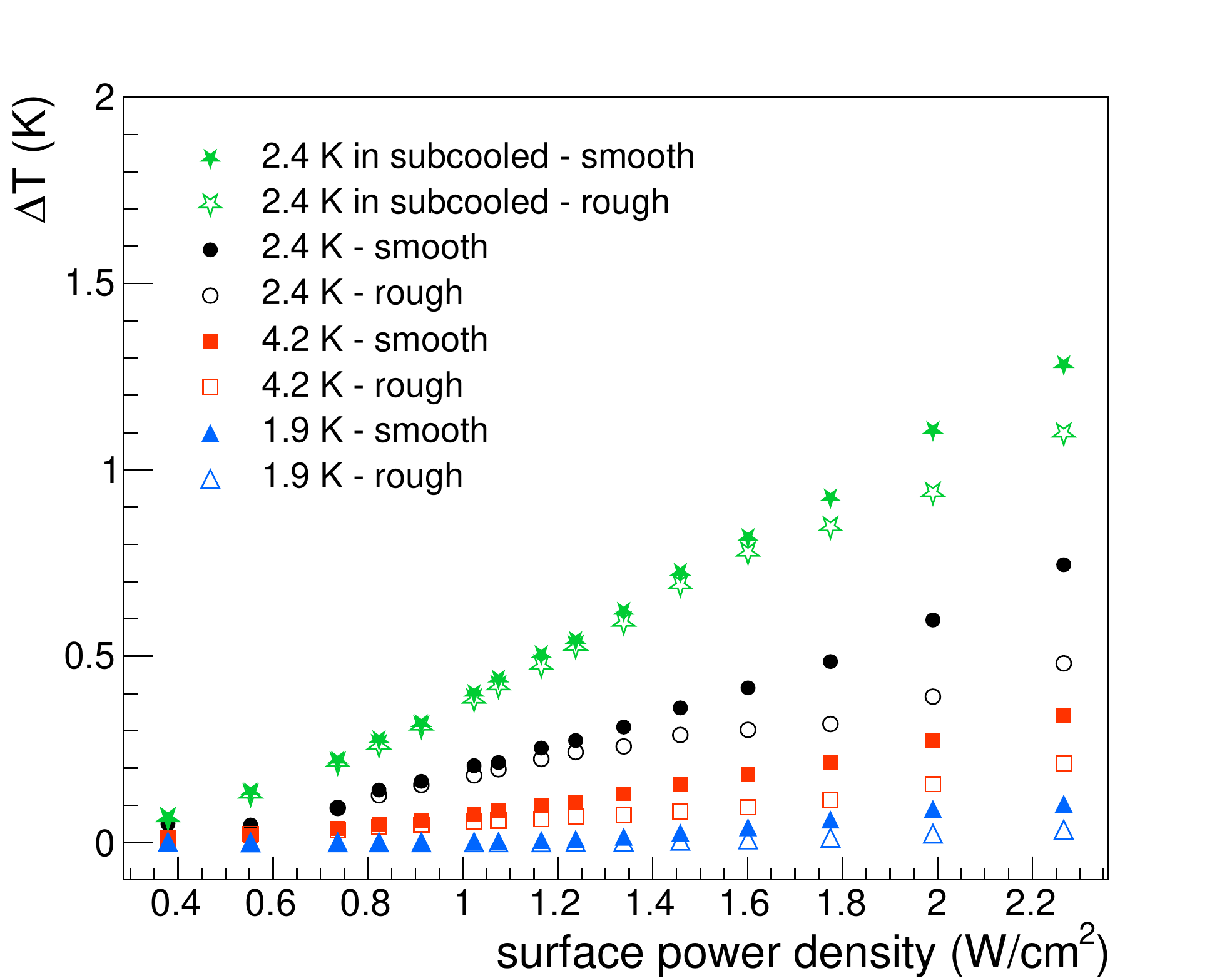}
   \caption{Temperature variations measured by the external thermometer at 1 cm from the heater on the smooth and rough Cu surfaces at 1.9 K (blue), 2.4 K (black), 4.2 K (red) at saturation pressure, and $\sim$2.4 K in subcooled He (green). Some statistical error bars are hidden by markers.}
   \label{fig:plot6}
\end{figure}

Figure \ref{fig:plot5} shows the temperature variations measured by the internal thermometer on the smooth and rough Cu surfaces at 1.9 K, 2.4 K, 4.2 K at saturation pressure, and $\sim$2.4 K in subcooled He, while figure \ref{fig:plot6} shows the temperature variations measured by the external thermometer on the same Cu surface and in the same He bath conditions previously mentioned.

As shown in figures \ref{fig:plot5} and \ref{fig:plot6}, the temperature values as a function of the surface power density depend on the roughness of the Cu surface. Indeed, temperature variations are generally higher on a smooth surface than on a rough surface. This is an important aspect to consider for superconducting thin film cavity applications. Indeed, with a rougher surface and a surface power density of 2 W/cm$^{2}$, the internal temperature of the Cu tube, measured 1 cm from the heater, decreases by $\sim$55\%, $\sim$40\%, $\sim$5\% and $\sim$15\% at 1.9 K, 4.2 K, 2.4 K at saturation pressure and $\sim$2.4 K in subcooled He, respectively. Similarly, the temperature measured by the external thermometer decreases by $\sim$75\%, $\sim$45\%, $\sim$35\%, and $\sim$15\% in the same He bath conditions mentioned above.

\section{\label{sec:level5}TEMPERATURE PROFILE ON COPPER SURFACES}

\begin{figure*}[!htb]
   \centering
   \includegraphics*[width=0.90\textwidth]{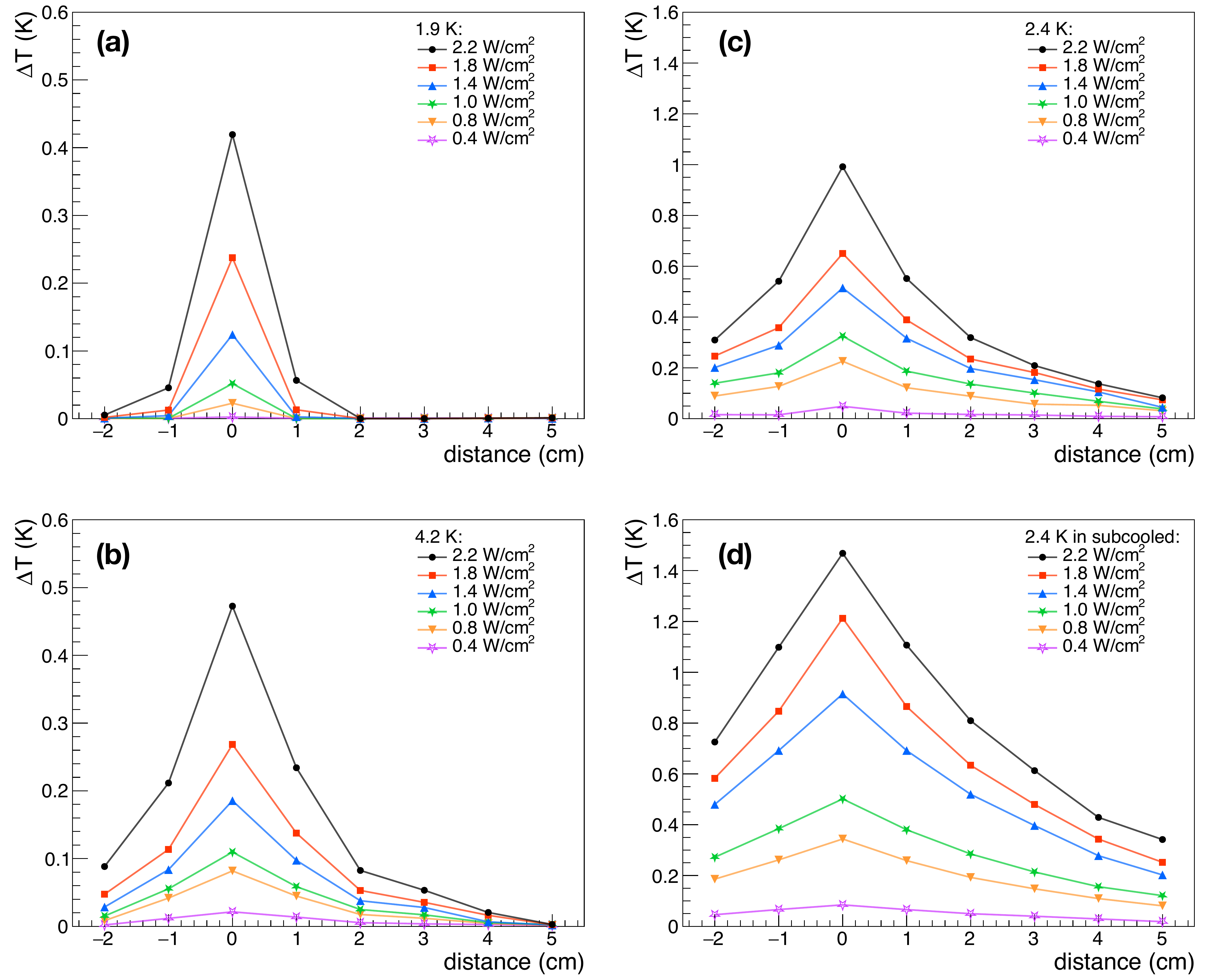}
   \caption{Temperature variations $\Delta T$ as a function of the distance from the heater. Temperature profiles are taken for values of surface power density ranging from 0.4 W/cm$^{2}$ to 2.2 W/cm$^{2}$ at 1.9 K (a), 4.2 K (b) and 2.4 K (c) in saturated vapor pressure, and $\sim$2.4 K in subcooled He (d). Statistical error bars are hidden by markers.}
   \label{fig:plot7e8e9e10}
\end{figure*}

Temperature profiles on Cu surfaces are affected by different parameters, like the thermal conductivity of Cu and He and the heat transfer from Cu to He. With the use of the thick film SMD resistor as a heater to simulate point-like heat loss in the Cu tube, temperature profiles on the outer surface of the tube are measured at different heat fluxes and He bath conditions. Temperature variations $\Delta T$ as a function of the distance from the heater are shown in figure \ref{fig:plot7e8e9e10} for different values of surface power density. Measurements have been carried out at varying conditions of He bath, in particular figures \ref{fig:plot7e8e9e10}a, \ref{fig:plot7e8e9e10}b, \ref{fig:plot7e8e9e10}c show data at 1.9 K, 4.2 K and 2.4 K in saturated vapor pressure while data at $\sim$2.4 K in subcooled He are shown in figure \ref{fig:plot7e8e9e10}d. The heater is placed at 0 cm, as shown in figure \ref{fig:plot7e8e9e10} where the peak of temperature profiles is located. Data are taken for values of surface power density ranging from $\sim$0.4 W/cm$^{2}$ to $\sim$2.3 W/cm$^{2}$ while the Cu surface is as received, as shown in figure \ref{fig:photo_surfaceflatandrough}a.

For a fixed value of heat flux, the temperature profile on the outer surface of the Cu tube in superfluid He at 1.9 K is generally lower than that at 4.2 K, as shown in figures \ref{fig:plot7e8e9e10}a and \ref{fig:plot7e8e9e10}b. At 1.0 W/cm$^{2}$, the temperature measured by the thermometer in correspondence to the heater is $\sim$50 mK, while it is $\sim$100 mK in normal liquid He at 4.2 K. It is important to observe that temperature profiles at 4.2 K have a higher width than those at 1.9 K. Indeed, the thermometers at 2 cm from the heater measure a temperature variation only at 4.2 K, while no variation is detected at 1.9 K for values of surface power density ranging from $\sim$0.4 W/cm$^{2}$ to $\sim$2.3 W/cm$^{2}$. At 2.4 K at saturation pressure, the temperature profiles are generally higher than those at 4.2 K by a factor of 3 to 4, whereas, in subcooled He, they are in general 3$-$6 times higher in comparison to those at 4.2 K, as shown in figure \ref{fig:plot7e8e9e10}d. The width of the temperature profiles at 2.4 K at saturation pressure and in subcooled He is much larger than that at 4.2 K, as examined in the next section.

\subsection{\label{sec:level6}Analysis of temperature profiles on copper surfaces}

A systematic study of the temperature profiles on Cu surfaces has been carried out to evaluate their height and width as the surface power density changes and the He bath conditions vary. The best interpolation of temperature profiles is obtained by the sum of two Gaussian functions with the same mean value $\mu$ as follows:
\begin{equation}\label{equation_fit}
\Delta T = p_{1} \; \exp \left(-\frac{(x - \mu)^{2}}{2 \sigma_{1}} \right) + p_{2} \; \exp \left(-\frac{(x - \mu)^{2}}{2 \sigma_{2}} \right)
\end{equation}
where $x$ is the distance from the heater, $p_{1}$ and $p_{2}$ are the height of the peaks of the first and second Gaussian curve, respectively, and $\sigma_{1}$ and $\sigma_{2}$ their width. As the heater is placed at 0 cm, the mean value $\mu$ of both Gaussian functions is assumed to be equal to 0 cm.

\begin{figure}[!htb]
   \centering
   \includegraphics*[width=0.65\columnwidth]{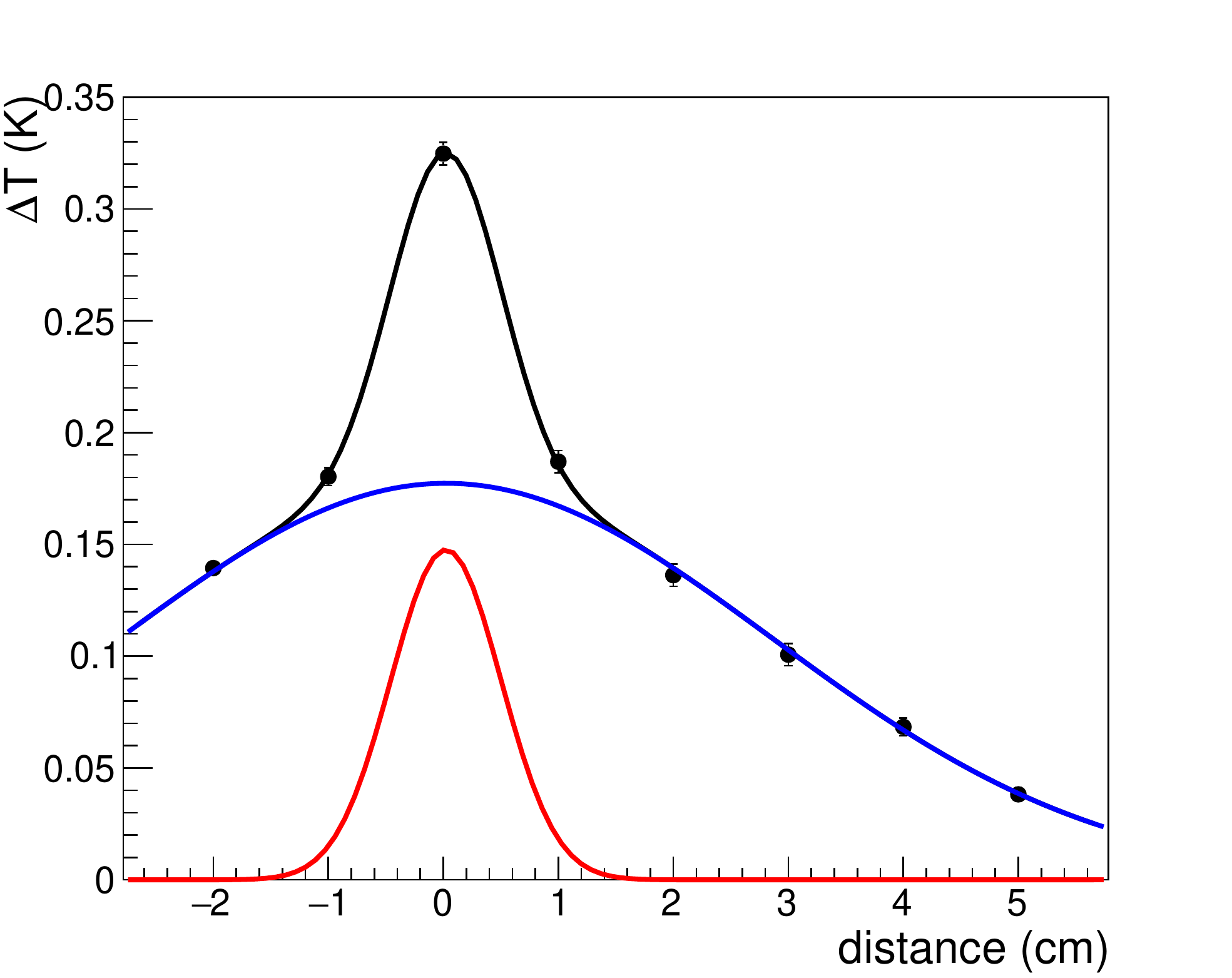} 
   \caption{Interpolation of temperature profile (in black) by the sum of two Gaussian functions (in red and blue), according to equation \ref{equation_fit}. The temperature profile is acquired at 2.4 K when the heater is fired at 1.0 W/cm$^{2}$. Some statistical error bars are hidden by markers.}
   \label{fig:plot11}
\end{figure}

Figure \ref{fig:plot11} shows the temperature profile at 2.4 K at saturation pressure when the heater is fired at 1.0 W/cm$^{2}$. The interpolation of temperature measurements by the sum of two Gaussian functions is indicated in black, whereas the single Gaussian curves are in red and blue. The red Gaussian curve, which is primarily influenced by the size of the heater, shows the maximum temperature variation in correspondence to the heat loss. Parameters of this curve can significantly vary with the usage of a different heater, especially the width $\sigma_{1}$ can change by using a larger or smaller heater. On the contrary, the Gaussian function in blue provides valuable information on the width of the temperature profile at a longer distance and outlines how the heat spreads on the Cu surface. Parameters of this second Gaussian function are more independent of heater dimensions and, consequently, are important for the design of temperature mapping systems. For instance, the density of thermometers and their sensitivity have to be designed and chosen depending on the values of $p_{2}$ and $\sigma_{2}$ in order to map all heat losses in the cavity with enough resolution.

\begin{figure}[!htb]
   \centering
   \includegraphics*[width=0.65\columnwidth]{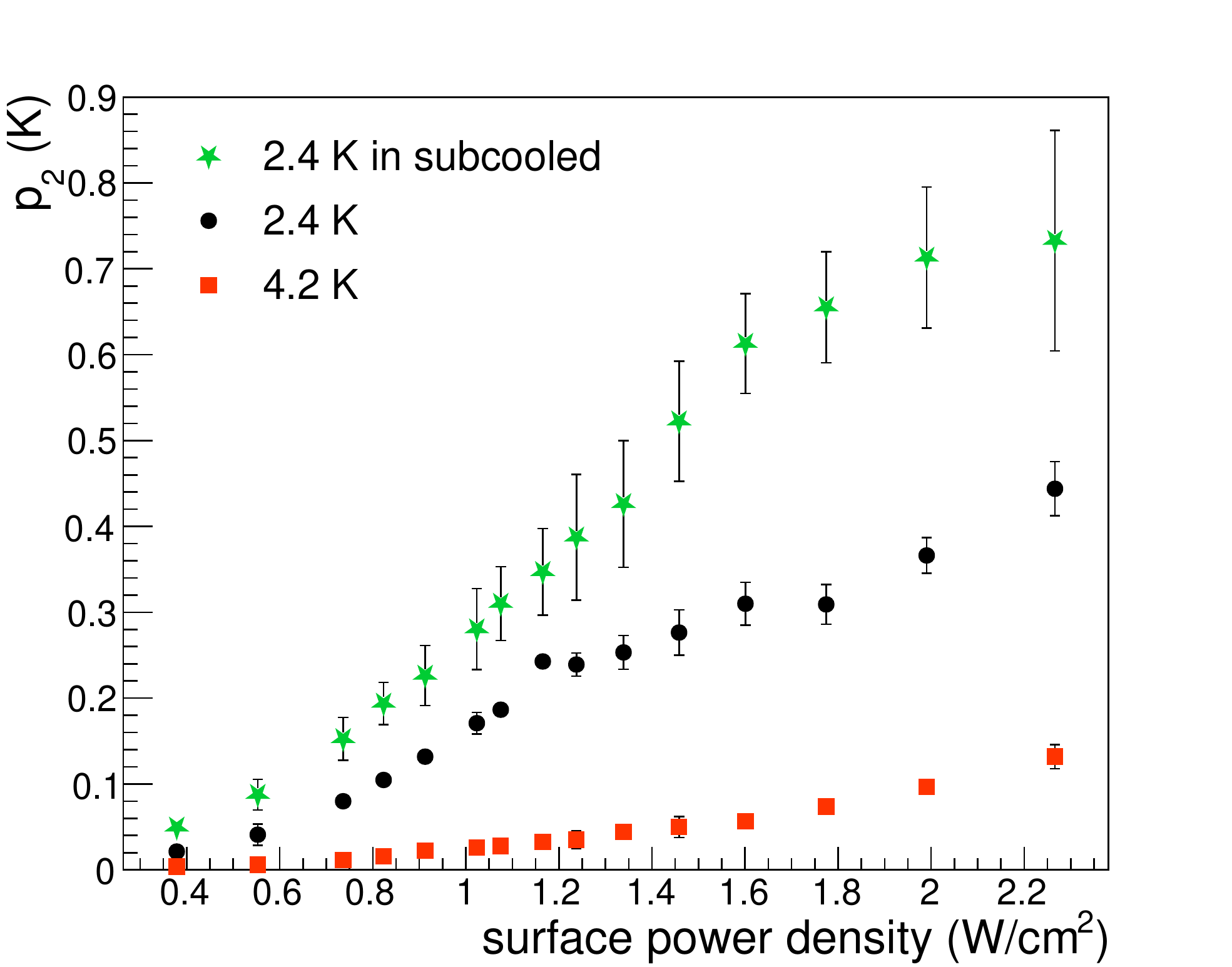}
   \caption{Values of $p_{2}$ from $\sim$0.4 W/cm$^{2}$ to $\sim$2.3 W/cm$^{2}$ at 4.2 K at saturation pressure (red), and at 2.4 K at saturation pressure (black) and subcooled He (green). Some statistical error bars are hidden by markers.}
   \label{fig:plot14}
\end{figure}

\begin{figure}[!htb]
   \centering
   \includegraphics*[width=0.65\columnwidth]{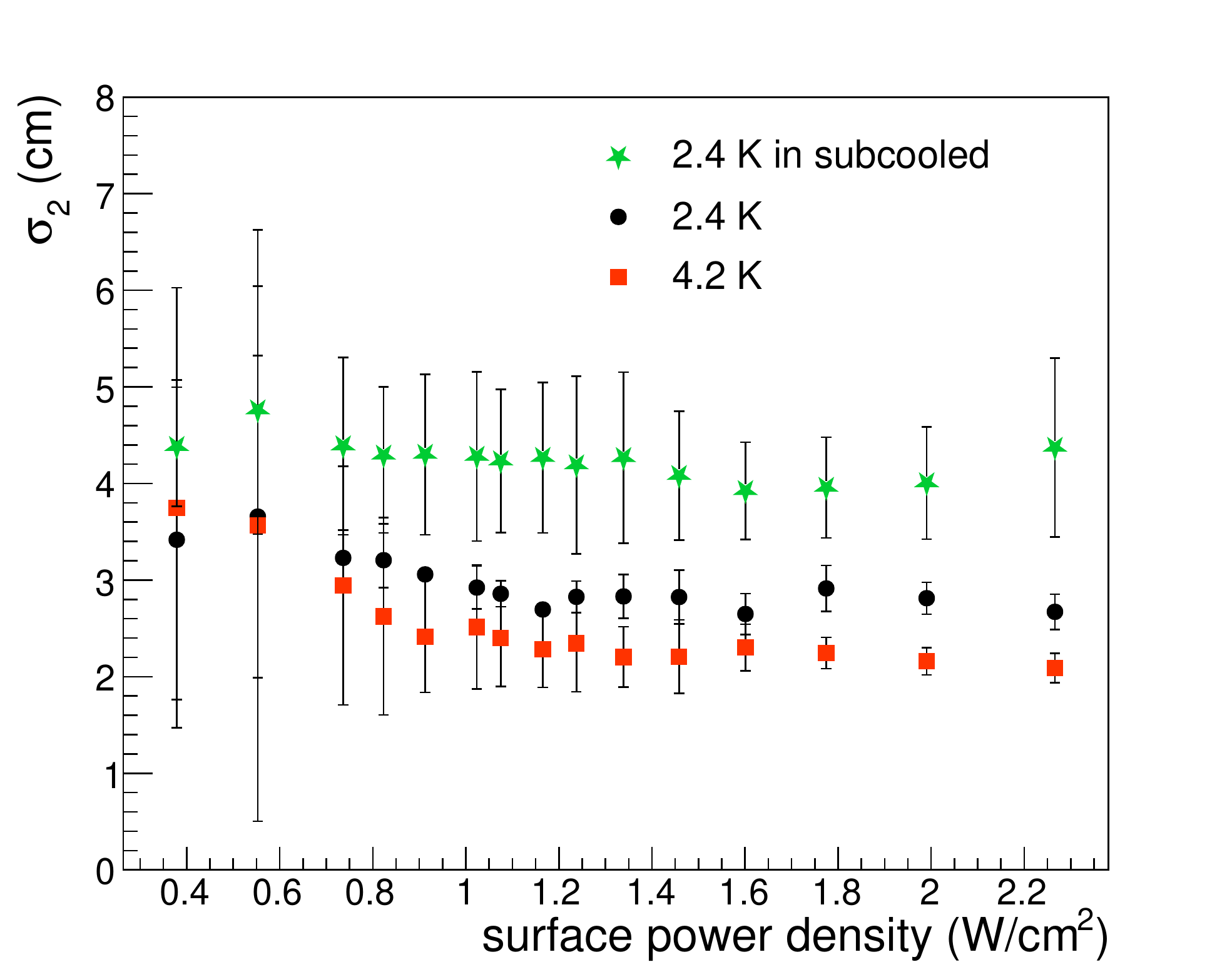} 
   \caption{Values of $\sigma_{2}$ from $\sim$0.4 W/cm$^{2}$ to $\sim$2.3 W/cm$^{2}$ at 4.2 K at saturation pressure (red), and at 2.4 K at saturation pressure (black) and subcooled He (green).}
   \label{fig:plot15}
\end{figure}

As $p_{1}$ and $\sigma_{1}$ are mainly correlated to the heater dimensions, we only report values of $p_{2}$ and $\sigma_{2}$ for different values of surface power density and temperatures of He bath. Values of $p_{2}$ and $\sigma_{2}$ are shown in figures \ref{fig:plot14} and \ref{fig:plot15}, respectively. These parameters are evaluated at 4.2 K and 2.4 K at saturation pressure and at $\sim$2.4 K in subcooled He. As expected, the height of the Gaussian curve $p_{2}$ increases progressively by increasing the surface power density. At 4.2 K, $p_{2}$ is $\sim$25 mK at $\sim$1 W/cm$^{2}$ and $\sim$100 mK at $\sim$2 W/cm$^{2}$. $p_{2}$ values at 2.4 K at saturation pressure are higher than those at 4.2 K. Indeed, $p_{2}$ at 2.4 K is $\sim$170 mK at $\sim$1 W/cm$^{2}$ and $\sim$360 mK at $\sim$2 W/cm$^{2}$. At $\sim$2.4 K in subcooled He, $p_{2}$ values are generally 60$-$70\% higher than those at 2.4 K at saturation pressure. Concerning the width of the Gaussian curve, the average value of $\sigma_{2}$ at 4.2 K is (2.21 $\pm$ 0.07) cm, whereas it rises to (2.92 $\pm$ 0.03) cm and (4.2 $\pm$ 0.2) cm at 2.4 K at saturation pressure and in subcooled He, respectively.

The analysis of $p_{2}$ and $\sigma_{2}$ cannot be considered reliable at 1.9 K due to the narrow width of most temperature profiles measured at this temperature of He bath, as shown in figure \ref{fig:plot7e8e9e10}a.

\subsection{\label{sec:level7}Efficiency of thermometers}

The efficiency of thermometers is defined as the ratio of the measured temperature rise to the expected temperature rise. With the experimental set-up described in section \ref{sec:level2}, we can measure the internal and external temperature of the Cu tube at 1 cm from the heater. In particular, one thermometer is glued to the inner surface of the Cu tube at 1 cm from the heater, whereas a second thermometer is pushed towards the outer surface of the Cu tube by a spring-loaded pin and measures the temperature variation in correspondence with the internal thermometer at the same distance from the heater. Due to the high thermal conductivity of the Cu surface and its low thickness (2 mm), we can assume that the temperature gradient along the thickness of the surface is negligible. In other words, the temperature on the outer surface of the Cu tube is in good approximation equal to that on the inner surface. The internal pressure of the Cu tube is lower than $\sim$10$^{-4}$ Pa, so it is reasonable to expect that the internal thermometer measures the exact temperature of the Cu surface at 1 cm from the heater. On the contrary, the external thermometer is immersed in the He bath, so the temperature rise measured by this thermometer is only a fraction of the temperature of the outer surface. Indeed, the external thermometer is heated on one side because it is in contact with the Cu surface and cooled by the He bath on the other side. Consequently, we can evaluate the efficiency of thermometers as the ratio of the temperature rise, measured by the external thermometer, to the temperature rise, measured by the internal thermometer.

\begin{figure}[!htb]
   \centering
   \includegraphics*[width=0.65\columnwidth]{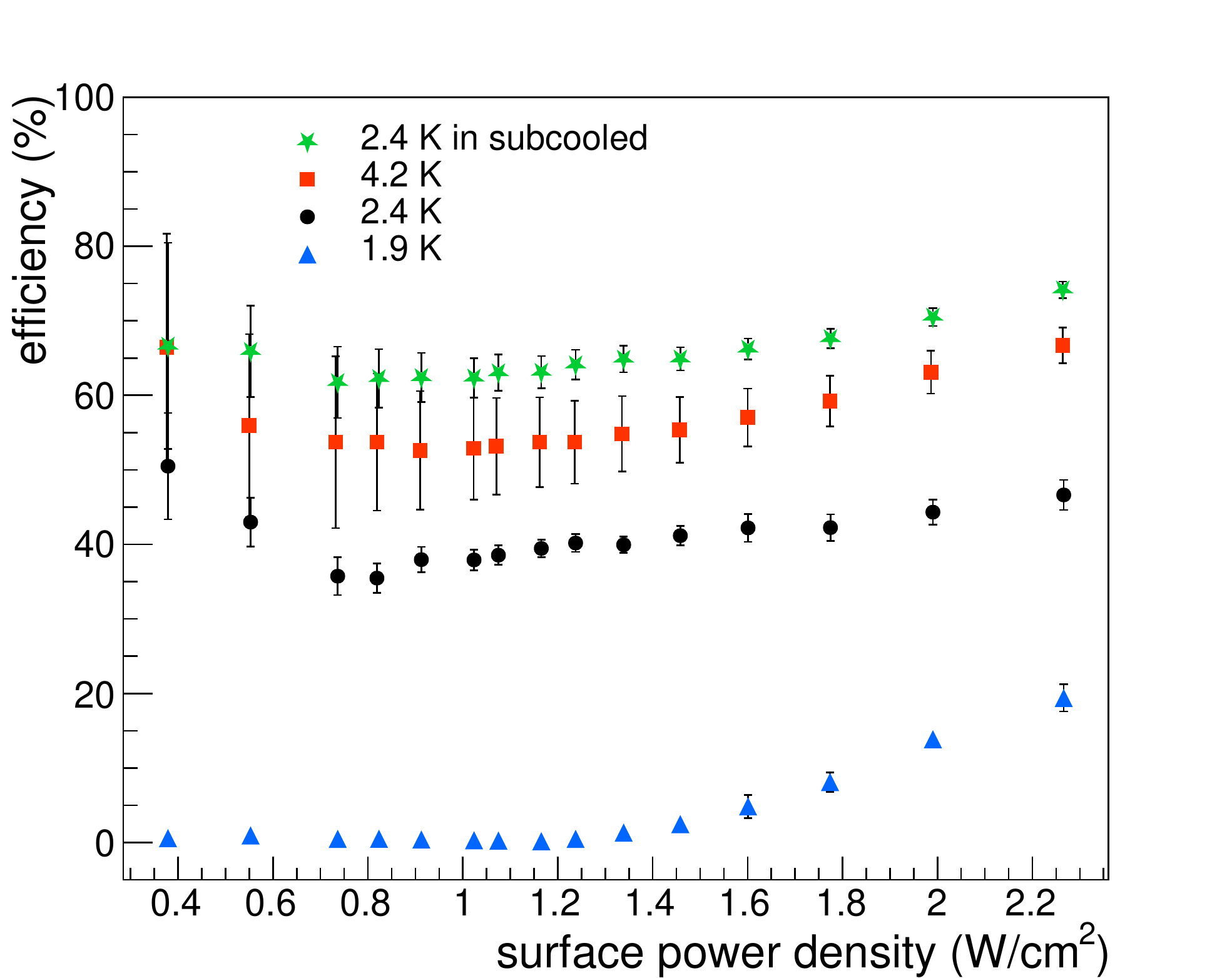} 
   \caption{Efficiency of thermometers from $\sim$0.4 W/cm$^{2}$ to $\sim$2.3 W/cm$^{2}$ at 1.9 K (blue), 2.4 K (black), 4.2 K (red) at saturation pressure, and $\sim$2.4 K in subcooled He (green). Some statistical error bars are hidden by markers.}
   \label{fig:plot16}
\end{figure}

Figure \ref{fig:plot16} shows the efficiency of thermometers at 1.9 K, 2.4 K, 4.2 K at saturation pressure, and $\sim$2.4 K in subcooled He. In all cases, the efficiency is not constant for values of surface power density ranging from $\sim$0.4 W/cm$^{2}$ to $\sim$2.3 W/cm$^{2}$. In the case of 1.9 K, the efficiency is close to 0\% for values of surface power density lower than $\sim$1.3 W/cm$^{2}$, while it increases linearly for values higher than $\sim$1.3 W/cm$^{2}$ until it reaches 20\% at $\sim$2.3 W/cm$^{2}$. At 4.2 K and 2.4 K at saturation pressure and in subcooled He, the efficiency of thermometers decreases from $\sim$0.4 W/cm$^{2}$ to $\sim$1 W/cm$^{2}$ where its minimum value is approximately reached. For values higher than $\sim$1 W/cm$^{2}$, the efficiency slightly increases in all three conditions of He bath. The average efficiency at 4.2 K is (59.8 $\pm$ 1.2)\% with a minimum value of $\sim$53\% at $\sim$1 W/cm$^{2}$ and a maximum value of $\sim$67\% at 2.3 W/cm$^{2}$. At 2.4 K at saturation pressure, the average efficiency is (40.1 $\pm$ 0.4)\%, while it is (67.7 $\pm$ 0.5)\% at the same temperature but in subcooled conditions. At 2.4 K at saturation pressure, the minimum and maximum efficiency values are $\sim$35\% and $\sim$51\%, respectively, whereas they rise to $\sim$62\% and $\sim$74\% in subcooled He.

\section{\label{sec:level8}TEMPERATURE PROFILE ON A SUPERCONDUCTING ELLIPTICAL CAVITY}

During a vertical cold test of a Nb/Cu 1.3 GHz elliptical cavity, temperature values of the external Cu surface have been acquired by ten thermometers along a line of longitude (meridian) of the cavity cell. Thermometers, described in section \ref{sec:level2}, are pushed by spring-loaded pins towards the outer surface of the cavity and Apiezon N grease is used to improve the thermal contact, as presented in section \ref{sec:level3}.

In a superconducting cavity, the power dissipated per unit area $P_{diss}$ at the RF surface is given by:
\begin{equation}
P_{diss} = \frac{1}{2} R_{s}(T) |\textbf{H}|^{2}
\end{equation}
where $R_{s}(T)$ is the surface resistance of the superconducting thin film, which depends on the temperature $T$ of the film, and $|\textbf{H}|$ is the magnitude of the RF magnetic field. If we assume that the heat transfer from the cavity to the He bath is uniform along the whole surface and the surface resistance does not change in the cell, the temperature profile along each meridian of the cavity cell is linearly proportional to $|\textbf{H}|^{2}$ in good approximation.

\begin{figure}[!htb]
   \centering
   \includegraphics*[width=0.65\columnwidth]{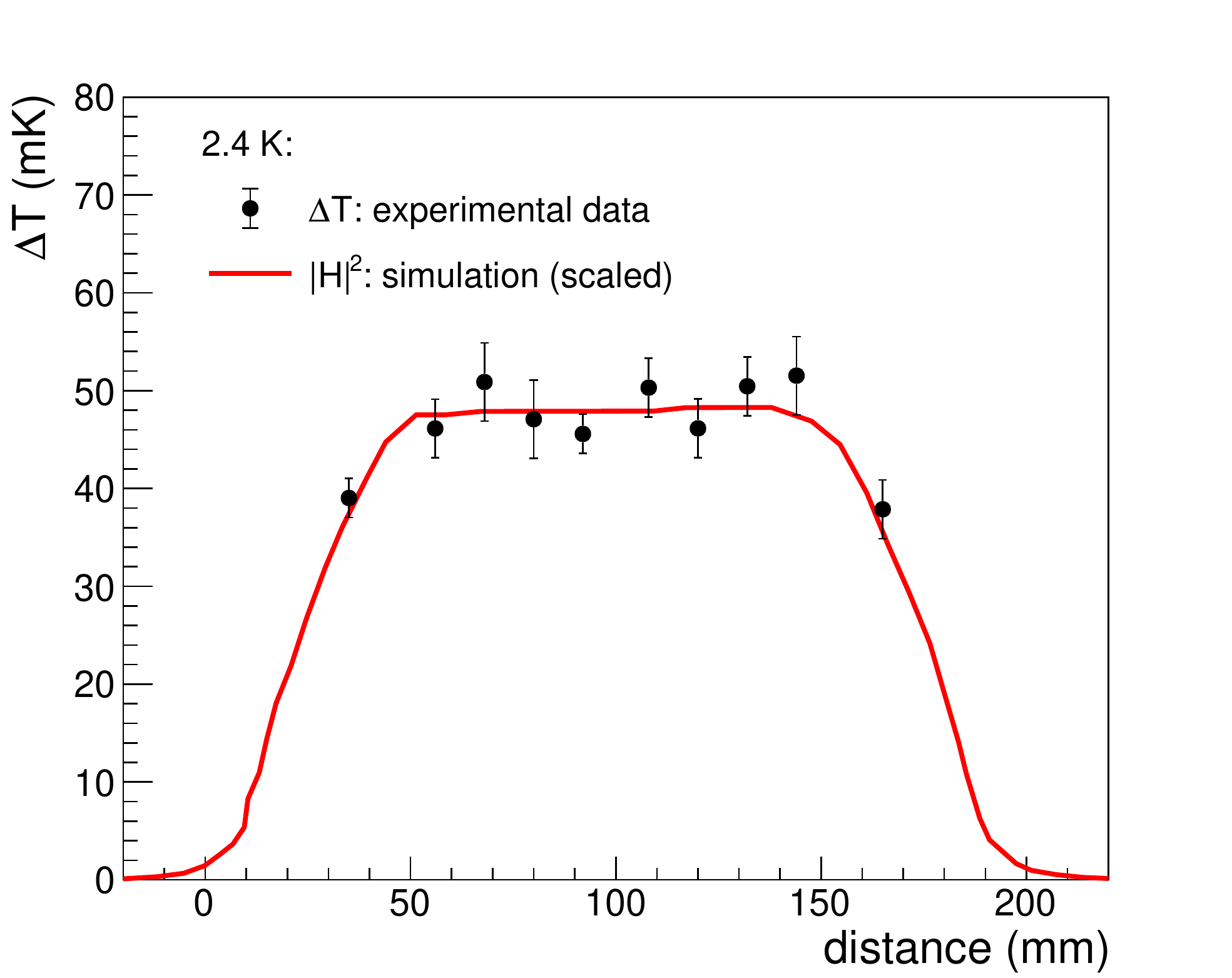} 
   \caption{Temperature profile measured by ten thermometers along a meridian of the cavity cell at 2.4 K at saturation pressure. The quality factor of the cavity is $\sim$10$^{10}$ at $\sim$6.2 MV/m. Along the meridian of the cavity cell, the temperature profile is compared with the profile of $|\textbf{H}|^{2}$, conveniently scaled.}
   \label{fig:plot17}
\end{figure}

Figure \ref{fig:plot17} shows the temperature profile measured by ten thermometers along a meridian of the cavity cell at 2.4 K at saturation pressure. The equator of the cavity is at 100 mm while the lower and upper cutoffs are at 0 and 200 mm, respectively. Data are acquired when the cavity is operated at $\sim$6.2 MV/m with a quality factor of $\sim$10$^{10}$. Thermometers are spaced $\sim$1 cm from each other. The profile of $|\textbf{H}|^{2}$ along the meridian of the cavity cell is simulated in COMSOL and conveniently scaled to overlap temperature measurements for easy comparison.

\begin{figure}[!htb]
   \centering
   \includegraphics*[width=0.65\columnwidth]{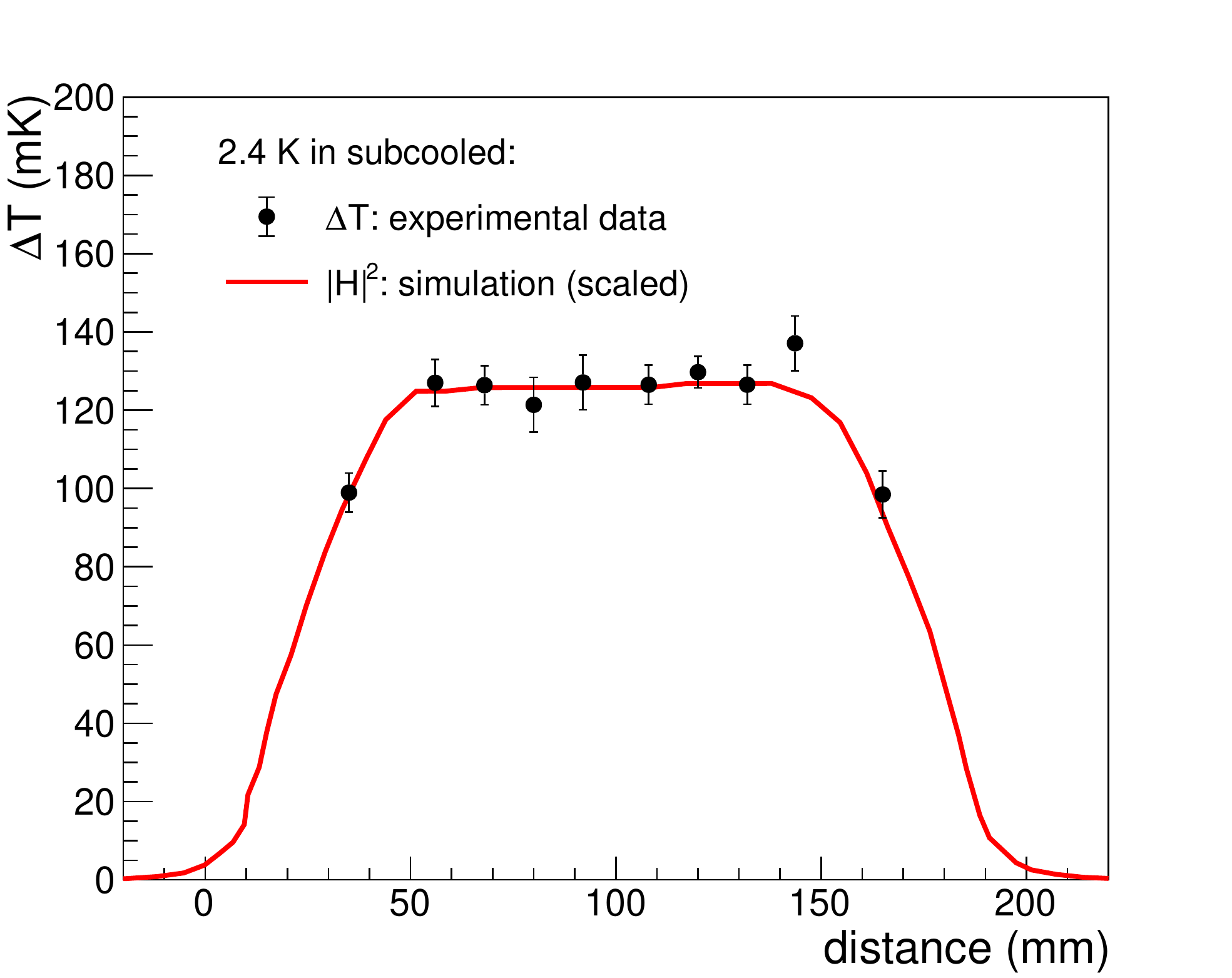} 
   \caption{Temperature profile measured by ten thermometers along a meridian of the cavity cell at $\sim$2.4 K in subcooled He. The quality factor of the cavity is $\sim$7$\cdot$10$^{9}$ at $\sim$6.8 MV/m. Along the meridian of the cavity cell, the temperature profile is compared with the profile of $|\textbf{H}|^{2}$, conveniently scaled.}
   \label{fig:plot18}
\end{figure}

In addition, figure \ref{fig:plot18} shows the temperature profile when the cavity is operated at $\sim$2.4 K in subcooled He at $\sim$6.8 MV/m with a quality factor of $\sim$7$\cdot$10$^{9}$.

The agreement between the expected temperature profile and temperature measurements is good in both cases by suggesting that the temperature mapping on thin film coated Cu cavities can be successfully carried out.

\section{\label{sec:level9}SUMMARY AND DISCUSSION}

\subsection{\label{sec:level10}Temperature mapping systems for thin film coated copper RF cavities}

This study aims at evaluating temperature rises on Cu surfaces with different roughness and thermal contacts in the presence of heat losses and for different temperatures of the He bath. The results of this study may help design temperature mapping systems for Nb/Cu cavities. Mapping heat losses on Cu cavities is challenging because of their high thermal conductivity at low temperatures. Several temperature mapping systems are currently in operation for bulk Nb cavities, whereas only one system was built in the '80s to map heat losses on Nb thin film on Cu cavities. This system only allowed for mapping the temperature of Nb/Cu SRF cavities in subcooled He \cite{piel1981cern}, which is not the standard operating condition of those cavities. Indeed, Nb/Cu cavities are usually operated in liquid He-I at saturated vapor pressure. Based on this study's encouraging outcomes, mapping the temperature on Cu coated SRF cavities may also be feasible in He-I at saturation pressure.

Allen-Bradley resistor thermometers are a valid solution to measure temperature variations on Cu surfaces at 1.9 K, 2.4 K, and 4.2 K at saturation pressure as well as at $\sim$2.4 K in subcooled He. The thermal contact between thermometers and Cu surfaces plays a crucial role in detecting the temperature rise of Cu surfaces in the presence of heat losses. If thermometers are just pushed towards Cu surfaces without any thermal paste in the interface, no temperature variations are detected in the presence of a point-like heat loss ranging from $\sim$0.4 W/cm$^{2}$ to $\sim$2.3 W/cm$^{2}$ in both He-II and He-I at saturation pressure. On the contrary, temperature variations can be detected at $\sim$2.4 K in subcooled He without any thermal paste. We found that Apiezon N grease improves thermal contact. Indeed, the temperature rises with this grease can be detected in all conditions of the He bath at saturation pressure. In the case of subcooled He at $\sim$2.4 K, temperature measurements with Apiezon N grease are generally higher by a factor of 2 than those without thermal paste between $\sim$0.4 W/cm$^{2}$ and $\sim$2.3 W/cm$^{2}$.

In He-I at saturation pressure, the ideal condition to detect heat losses on Cu surfaces is at 2.4 K because the thermal conductivity of Cu and He-I at this temperature is lower than that at 4.2 K, as discussed in section \ref{sec:level2}. This implies that the temperature profiles at 2.4 K on Cu surfaces are higher and broader than those at 4.2 K in the presence of a point-like heat loss, as demonstrated in section \ref{sec:level5}. In particular, parameters $p_{2}$ and $\sigma_{2}$, which correspond to the height and the width of temperature profiles as defined in section \ref{sec:level6}, are essential to determine the spacing of thermometers for superconducting thin film applications. In addition, the sensitivity of Allen-Bradley resistors at 2.4 K is higher than that at 4.2 K.

At $\sim$2.4 K in subcooled He, we found that the temperature profiles on Cu surfaces are much higher and broader than that at 2.4 K at saturation pressure, as shown in figure \ref{fig:plot7e8e9e10} and already described by H. Piel in the '80s \cite{piel1980diagnostic, piel1981cern}. Indeed, the thermal properties of Cu as well as the sensitivity of thermometers do not change, but the overpressure in the He bath suppresses the formation of gaseous He bubbles so that the heat transport by the nucleate boiling regime is inhibited. As a result, the Cu surfaces are mainly cooled through convection cooling, which is less efficient than nucleate boiling, leading to an increase in temperature near hotspots \cite{padamsee2020history}.

The measurement of temperature increase on Cu surfaces in He-II is more challenging than that in He-I. This is because even though the thermal conductivity of Cu in He-II is reduced in comparison to that in He-I and the sensitivity of Allen-Bradley resistors is higher than that in He-I, the cooling capability of He-II is much higher than that of He-I, which makes the detection of temperature rises challenging.

In section \ref{sec:level8}, we showed temperature measurements on the outer surface of a Nb/Cu elliptical cavity with a resonance frequency of 1.3 GHz. The temperature is measured by ten thermometers spaced by $\sim$1 cm each other and placed along a meridian of the cavity cell. The expected temperature profile is well reproduced by experimental measurements in both the equator of the cavity cell and its edges, where the temperature is expected to be lower than that at the equator. A good agreement between experimental data and simulation results is observed at 2.4 K at saturation pressure and in subcooled He. These promising results imply that the detection and the localization of heat losses in thin film coated Cu cavities is possible by using Allen-Bradley resistors with Apiezon N grease at 2.4 K in subcooled He, but also at saturated vapor pressure where thin film coated Cu cavities are usually operated.

\subsection{\label{sec:level11}Heat dissipation in thin film coated copper RF cavities}

According to our results, the roughness of the surface plays a crucial role in the heat transfer from Cu surfaces to He bath, as also reported in the literature \cite{smith1969review, bald1976nucleate, schmidt1981review}. Therefore, this aspect needs to be carefully considered in superconducting thin film cavity applications. Indeed, increasing the roughness of the outer surface of these cavities may improve the heat transfer into the He bath and, in turn, their RF performance.

The density of nucleation sites is higher in rough Cu surfaces, as examined in section \ref{sec:level4}. The activation of nucleation sites is fundamental to enhancing the heat transfer from the surface to the He bath. Indeed, the activation of nucleation sites implies the onset of He bubble formation on the surface and, in turn, the transition from convection cooling to nucleate boiling regime, which is more effective than just convection cooling \cite{schmidt1981review}. 

Our results, shown in figures \ref{fig:plot17} and \ref{fig:plot18}, are consistent with data in the literature, where several authors observed that the heat transfer from Cu surfaces to He-I is generally affected by the surface roughness \cite{smith1969review, bald1976nucleate, schmidt1981review}. This is also observed in He-II in many cases \cite{amrit2000heat, amrit2013kapitza}. 

The present findings have further strengthened our hypothesis that cooling thin film coated Cu cavities by nucleation boiling could improve their RF performance. For a constant heat flux, the superconducting thin films in SRF cavities with an enhanced heat transfer into the He bath would remain at a lower temperature compared to those in cavities with a worse heat transfer. According to \cite{halbritter1970comparison}, when the applied RF frequency inside the cavity is much lower than the energy gap of the superconductor and the temperature is below half of the superconducting transition temperature, the surface resistance $R$ of a superconductor depends on the temperature $T$ as follows: 
\begin{equation}
    R(T) = \frac{A_{0}}{T} \exp \left(-\frac{\Delta_{0}}{k_{B}T}\right)
\end{equation}
where $A_{0}$ depends on material parameters and RF frequency, $\Delta_{0}$ is the superconducting gap, and $k_{B}$ is the Boltzmann constant. Therefore, limiting the temperature increase of the superconducting thin film in RF cavities by improving their heat dissipation implies the improvement of the RF performance. This is particularly important for cavities operated in liquid He-I at temperatures between 4.0 K and 4.5 K, like superconducting Nb thin film onto Cu cavities. For example, in the case of a Nb/Cu 1.3 GHz elliptical cavity, the variation of surface resistance is generally $\sim$250 n$\Omega$/K at 4.2 K, whereas it is only $\sim$7 n$\Omega$/K at 1.9 K. In other words, if the temperature rise of the Nb thin film is of 1 mK during cavity operations, its surface resistance is increased by $\sim$7 p$\Omega$ in He-II, but the increment of resistance is more than 35 times higher in He-I at 4.2 K where the cavity is usually operated.

Increasing the roughness of the outer surface of thin film coated Cu cavities might improve their RF performance. If the cooling by nucleate boiling regime is induced in large areas of thin film coated Cu cavities by increasing their external roughness, the heat exchange from the cavities to the He bath is improved. The result is that the RF performance is improved in cavities with a more efficient heat transfer into the He bath. In addition, increasing the external roughness of thin film cavities may mitigate the degradation of RF performance in the presence of hotspots in the internal film, which might be present in large coated areas.

\section{\label{sec:level12}CONCLUSION}


This paper reports how temperature variations on Cu surfaces can be satisfactorily measured and how the heat dissipation in thin film coated Cu cavities could be enhanced. These observations could have several implications for designing superconducting thin film cavity applications at liquid He temperatures, like temperature mapping systems for Nb thin film coated Cu cavities. 

The ideal condition to detect heat losses on Cu surfaces is slightly above the lambda point of He because the thermal conductivity of Cu and liquid He is lower than that at higher temperatures. Measuring temperature increases on Cu surfaces at $\sim$2.4 K in subcooled He is found to be advantageous. However, this condition of He bath is not stable in time since the temperature slowly drifts to higher values and, in addition, thin film cavities are usually operated at saturation pressure. Therefore, a valid alternative is to carry out temperature measurements at 2.4 K at saturation pressure where the temperature profiles are higher and broader than those measured at 4.2 K and 1.9 K, but slightly lower and narrower than those at $\sim$2.4 K in subcooled He. This is confirmed by measuring the temperature profile along a meridian of a 1.3 GHz elliptical cavity cell at 2.4 K both at saturation pressure and in subcooled He. A good agreement between experimental measurements and simulation results is obtained in both cases.

This study has gone some way toward enhancing our understanding of heat dissipation in thin film coated Cu cavities. According to the results of this study, increasing the external roughness of thin film coated Cu cavities may improve their RF performance. Indeed, the roughness of Cu surfaces plays a crucial role in the heat transfer from Cu to He bath, as already examined in this work. This is an aspect that should be carefully taken into account for superconducting thin film cavity applications. For a fixed heat flux, a rough surface in Cu is generally colder than a smooth surface. Therefore, the enhancement of heat exchange into the He bath may imply a reduced temperature increase of the superconducting thin film in RF cavities and, in turn, improve the RF performance. The present findings suggest that engineering the outer surface of thin film cavities by increasing the roughness of the Cu substrate might improve their RF performance.

\section*{\label{sec:level13}ACKNOWLEDGEMENTS}
We gratefully acknowledge the contribution of our colleagues Marco Chiodini, Pablo Vidal Garcia, Lorena Vega Cid, Laura Bergamaschi, and Louis Boyer for their support in the project, preparation, and testing. Our special thanks go to Akira Miyazaki and Torsten Koettig for the helpful discussion. We warmly thank Agostino Vacca and Florence Crochon for their invaluable help.

\section*{\label{sec:level13}DATA AVAILABILITY}
The data that support the findings of this study are available from the corresponding
author upon reasonable request.

\nocite{*}

\bibliography{apssamp}

\begin{thebibliography}{35}%
\makeatletter
\providecommand \@ifxundefined [1]{%
 \@ifx{#1\undefined}
}%
\providecommand \@ifnum [1]{%
 \ifnum #1\expandafter \@firstoftwo
 \else \expandafter \@secondoftwo
 \fi
}%
\providecommand \@ifx [1]{%
 \ifx #1\expandafter \@firstoftwo
 \else \expandafter \@secondoftwo
 \fi
}%
\providecommand \natexlab [1]{#1}%
\providecommand \enquote  [1]{``#1''}%
\providecommand \bibnamefont  [1]{#1}%
\providecommand \bibfnamefont [1]{#1}%
\providecommand \citenamefont [1]{#1}%
\providecommand \href@noop [0]{\@secondoftwo}%
\providecommand \href [0]{\begingroup \@sanitize@url \@href}%
\providecommand \@href[1]{\@@startlink{#1}\@@href}%
\providecommand \@@href[1]{\endgroup#1\@@endlink}%
\providecommand \@sanitize@url [0]{\catcode `\\12\catcode `\$12\catcode
  `\&12\catcode `\#12\catcode `\^12\catcode `\_12\catcode `\%12\relax}%
\providecommand \@@startlink[1]{}%
\providecommand \@@endlink[0]{}%
\providecommand \url  [0]{\begingroup\@sanitize@url \@url }%
\providecommand \@url [1]{\endgroup\@href {#1}{\urlprefix }}%
\providecommand \urlprefix  [0]{URL }%
\providecommand \Eprint [0]{\href }%
\providecommand \doibase [0]{https://doi.org/}%
\providecommand \selectlanguage [0]{\@gobble}%
\providecommand \bibinfo  [0]{\@secondoftwo}%
\providecommand \bibfield  [0]{\@secondoftwo}%
\providecommand \translation [1]{[#1]}%
\providecommand \BibitemOpen [0]{}%
\providecommand \bibitemStop [0]{}%
\providecommand \bibitemNoStop [0]{.\EOS\space}%
\providecommand \EOS [0]{\spacefactor3000\relax}%
\providecommand \BibitemShut  [1]{\csname bibitem#1\endcsname}%
\let\auto@bib@innerbib\@empty
\bibitem [{\citenamefont {Padamsee}\ \emph {et~al.}(2009)\citenamefont
  {Padamsee}, \citenamefont {Knobloch},\ and\ \citenamefont
  {Hays}}]{padamsee2}%
  \BibitemOpen
  \bibfield  {author} {\bibinfo {author} {\bibfnamefont {H.}~\bibnamefont
  {Padamsee}}, \bibinfo {author} {\bibfnamefont {J.}~\bibnamefont {Knobloch}},\
  and\ \bibinfo {author} {\bibfnamefont {T.}~\bibnamefont {Hays}},\ }\bibfield
  {title} {\bibinfo {title} {Cavity fabrication and preparation},\ }in\
  \href@noop {} {\emph {\bibinfo {booktitle} {{RF} Superconductivity for
  Accelerators}}}\ (\bibinfo  {publisher} {Wiley},\ \bibinfo {year}
  {2009})\BibitemShut {NoStop}%
\bibitem [{\citenamefont {Benvenuti}\ \emph {et~al.}(1984)\citenamefont
  {Benvenuti}, \citenamefont {Circelli},\ and\ \citenamefont
  {Hauer}}]{benvenuti1984niobium}%
  \BibitemOpen
  \bibfield  {author} {\bibinfo {author} {\bibfnamefont {C.}~\bibnamefont
  {Benvenuti}}, \bibinfo {author} {\bibfnamefont {N.}~\bibnamefont
  {Circelli}},\ and\ \bibinfo {author} {\bibfnamefont {M.}~\bibnamefont
  {Hauer}},\ }\bibfield  {title} {\bibinfo {title} {Niobium films for
  superconducting accelerating cavities},\ }\href@noop {} {\bibfield  {journal}
  {\bibinfo  {journal} {Applied Physics Letters}\ }\textbf {\bibinfo {volume}
  {45}},\ \bibinfo {pages} {583} (\bibinfo {year} {1984})}\BibitemShut
  {NoStop}%
\bibitem [{\citenamefont {Padamsee}(2009)}]{padamsee}%
  \BibitemOpen
  \bibfield  {author} {\bibinfo {author} {\bibfnamefont {H.}~\bibnamefont
  {Padamsee}},\ }\bibfield  {title} {\bibinfo {title} {Cavity fabrication
  advances},\ }in\ \href@noop {} {\emph {\bibinfo {booktitle} {{RF}
  Superconductivity: Science, Technology and Applications}}}\ (\bibinfo
  {publisher} {Wiley},\ \bibinfo {year} {2009})\BibitemShut {NoStop}%
\bibitem [{\citenamefont {Halbritter}(1974)}]{halbritter1974surface}%
  \BibitemOpen
  \bibfield  {author} {\bibinfo {author} {\bibfnamefont {J.}~\bibnamefont
  {Halbritter}},\ }\bibfield  {title} {\bibinfo {title} {On surface resistance
  of superconductors},\ }\href@noop {} {\bibfield  {journal} {\bibinfo
  {journal} {Zeitschrift fur Physik}\ }\textbf {\bibinfo {volume} {266}},\
  \bibinfo {pages} {209} (\bibinfo {year} {1974})}\BibitemShut {NoStop}%
\bibitem [{\citenamefont {Aull}\ \emph {et~al.}(2015)\citenamefont {Aull}, ,
  \citenamefont {Junginger}, \citenamefont {Knobloch}, \citenamefont {Sublet},
  \citenamefont {Valente-Feliciano}, \citenamefont {Venturini~Delsolaro},\ and\
  \citenamefont {Zhang}}]{aull}%
  \BibitemOpen
  \bibfield  {author} {\bibinfo {author} {\bibfnamefont {S.}~\bibnamefont
  {Aull}}, , \bibinfo {author} {\bibfnamefont {T.}~\bibnamefont {Junginger}},
  \bibinfo {author} {\bibfnamefont {J.}~\bibnamefont {Knobloch}}, \bibinfo
  {author} {\bibfnamefont {A.}~\bibnamefont {Sublet}}, \bibinfo {author}
  {\bibfnamefont {A.-M.}\ \bibnamefont {Valente-Feliciano}}, \bibinfo {author}
  {\bibfnamefont {W.}~\bibnamefont {Venturini~Delsolaro}},\ and\ \bibinfo
  {author} {\bibfnamefont {P.}~\bibnamefont {Zhang}},\ }\bibfield  {title}
  {\bibinfo {title} {On the understanding of {Q}-slope of niobium thin films},\
  }\href@noop {} {\bibfield  {journal} {\bibinfo  {journal} {17th International
  Conference on RF Superconductivity}\ }\textbf {\bibinfo {volume} {TUBA03}}
  (\bibinfo {year} {2015})}\BibitemShut {NoStop}%
\bibitem [{\citenamefont {Miyazaki}\ and\ \citenamefont
  {Delsolaro}(2019)}]{miyazaki2019two}%
  \BibitemOpen
  \bibfield  {author} {\bibinfo {author} {\bibfnamefont {A.}~\bibnamefont
  {Miyazaki}}\ and\ \bibinfo {author} {\bibfnamefont {W.~V.}\ \bibnamefont
  {Delsolaro}},\ }\bibfield  {title} {\bibinfo {title} {Two different origins
  of the {Q}-slope problem in superconducting niobium film cavities for a heavy
  ion accelerator at cern},\ }\href@noop {} {\bibfield  {journal} {\bibinfo
  {journal} {Physical Review Accelerators and Beams}\ }\textbf {\bibinfo
  {volume} {22}},\ \bibinfo {pages} {073101} (\bibinfo {year}
  {2019})}\BibitemShut {NoStop}%
\bibitem [{\citenamefont {Bruning}\ \emph {et~al.}(2004)\citenamefont
  {Bruning}, \citenamefont {Collier}, \citenamefont {Lebrun}, \citenamefont
  {Myers}, \citenamefont {Ostojic}, \citenamefont {Poole},\ and\ \citenamefont
  {Proudlock}}]{Bruning:782076}%
  \BibitemOpen
  \bibfield  {author} {\bibinfo {author} {\bibfnamefont {O.~S.}\ \bibnamefont
  {Bruning}}, \bibinfo {author} {\bibfnamefont {P.}~\bibnamefont {Collier}},
  \bibinfo {author} {\bibfnamefont {P.}~\bibnamefont {Lebrun}}, \bibinfo
  {author} {\bibfnamefont {S.}~\bibnamefont {Myers}}, \bibinfo {author}
  {\bibfnamefont {R.}~\bibnamefont {Ostojic}}, \bibinfo {author} {\bibfnamefont
  {J.}~\bibnamefont {Poole}},\ and\ \bibinfo {author} {\bibfnamefont
  {P.}~\bibnamefont {Proudlock}},\ }\href@noop {} {\emph {\bibinfo {title} {LHC
  Design Report}}}\ (\bibinfo  {publisher} {CERN},\ \bibinfo {year}
  {2004})\BibitemShut {NoStop}%
\bibitem [{\citenamefont {Benvenuti}\ \emph {et~al.}(1991)\citenamefont
  {Benvenuti}, \citenamefont {Bloess}, \citenamefont {Hilleret}, \citenamefont
  {Bernard}, \citenamefont {Tuckmantel}, \citenamefont {Chiaveri},
  \citenamefont {Cavallari}, \citenamefont {Habel},\ and\ \citenamefont
  {Weingarten}}]{benvenuti1991superconducting}%
  \BibitemOpen
  \bibfield  {author} {\bibinfo {author} {\bibfnamefont {C.}~\bibnamefont
  {Benvenuti}}, \bibinfo {author} {\bibfnamefont {D.}~\bibnamefont {Bloess}},
  \bibinfo {author} {\bibfnamefont {N.}~\bibnamefont {Hilleret}}, \bibinfo
  {author} {\bibfnamefont {P.}~\bibnamefont {Bernard}}, \bibinfo {author}
  {\bibfnamefont {J.}~\bibnamefont {Tuckmantel}}, \bibinfo {author}
  {\bibfnamefont {E.}~\bibnamefont {Chiaveri}}, \bibinfo {author}
  {\bibfnamefont {G.}~\bibnamefont {Cavallari}}, \bibinfo {author}
  {\bibfnamefont {E.}~\bibnamefont {Habel}},\ and\ \bibinfo {author}
  {\bibfnamefont {W.}~\bibnamefont {Weingarten}},\ }\href@noop {} {\emph
  {\bibinfo {title} {Superconducting niobium sputter-coated copper cavity
  modules for the LEP energy upgrade}}},\ \bibinfo {type} {Tech. Rep.}\
  (\bibinfo {year} {1991})\BibitemShut {NoStop}%
\bibitem [{\citenamefont {Chiaveri}(1995)}]{chiaveri1995production}%
  \BibitemOpen
  \bibfield  {author} {\bibinfo {author} {\bibfnamefont {E.}~\bibnamefont
  {Chiaveri}},\ }\bibfield  {title} {\bibinfo {title} {Production by industry
  of a large number of superconducting cavities: Status and future},\
  }\href@noop {} {\bibfield  {journal} {\bibinfo  {journal} {Part. Accel.}\
  }\textbf {\bibinfo {volume} {53}},\ \bibinfo {pages} {253} (\bibinfo {year}
  {1995})}\BibitemShut {NoStop}%
\bibitem [{\citenamefont {Schirm}\ \emph {et~al.}(1995)\citenamefont {Schirm},
  \citenamefont {Thony}, \citenamefont {Weingarten},\ and\ \citenamefont
  {Chiaveri}}]{schirm1995analysis}%
  \BibitemOpen
  \bibfield  {author} {\bibinfo {author} {\bibfnamefont {K.}~\bibnamefont
  {Schirm}}, \bibinfo {author} {\bibfnamefont {B.}~\bibnamefont {Thony}},
  \bibinfo {author} {\bibfnamefont {W.}~\bibnamefont {Weingarten}},\ and\
  \bibinfo {author} {\bibfnamefont {E.}~\bibnamefont {Chiaveri}},\ }\href@noop
  {} {\emph {\bibinfo {title} {Analysis of the coating success rate in the
  series production of Nb/Cu superconducting RF cavities for LEP2}}},\ \bibinfo
  {type} {Tech. Rep.}\ (\bibinfo  {institution} {CM-P00062549},\ \bibinfo
  {year} {1995})\BibitemShut {NoStop}%
\bibitem [{\citenamefont {Chiaveri}(1999)}]{chiaveri1999cern}%
  \BibitemOpen
  \bibfield  {author} {\bibinfo {author} {\bibfnamefont {E.}~\bibnamefont
  {Chiaveri}},\ }\href@noop {} {\emph {\bibinfo {title} {The CERN Nb/Cu
  Programme for the LHC and Reduced-b Superconducting Cavities}}},\ \bibinfo
  {type} {Tech. Rep.}\ (\bibinfo {year} {1999})\BibitemShut {NoStop}%
\bibitem [{\citenamefont {Venturini~Delsolaro}\ \emph
  {et~al.}(2013)\citenamefont {Venturini~Delsolaro}, \citenamefont {Jecklin},
  \citenamefont {Kadi}, \citenamefont {Mondino}, \citenamefont {Sublet},
  \citenamefont {Delaup}, \citenamefont {Palmieri}, \citenamefont {Calatroni},
  \citenamefont {Stark}, \citenamefont {Therasse} \emph
  {et~al.}}]{venturini2013nb}%
  \BibitemOpen
  \bibfield  {author} {\bibinfo {author} {\bibfnamefont {W.}~\bibnamefont
  {Venturini~Delsolaro}}, \bibinfo {author} {\bibfnamefont {N.}~\bibnamefont
  {Jecklin}}, \bibinfo {author} {\bibfnamefont {Y.}~\bibnamefont {Kadi}},
  \bibinfo {author} {\bibfnamefont {I.}~\bibnamefont {Mondino}}, \bibinfo
  {author} {\bibfnamefont {A.}~\bibnamefont {Sublet}}, \bibinfo {author}
  {\bibfnamefont {B.}~\bibnamefont {Delaup}}, \bibinfo {author} {\bibfnamefont
  {V.}~\bibnamefont {Palmieri}}, \bibinfo {author} {\bibfnamefont
  {S.}~\bibnamefont {Calatroni}}, \bibinfo {author} {\bibfnamefont
  {S.}~\bibnamefont {Stark}}, \bibinfo {author} {\bibfnamefont
  {M.}~\bibnamefont {Therasse}}, \emph {et~al.},\ }\bibfield  {title} {\bibinfo
  {title} {Nb sputtered quarter wave resonators for the {HIE-ISOLDE}},\
  }\href@noop {} {\bibfield  {journal} {\bibinfo  {journal} {16th International
  Conference on RF Superconductivity}\ } (\bibinfo {year} {2013})}\BibitemShut
  {NoStop}%
\bibitem [{\citenamefont {Benvenuti}\ \emph {et~al.}(1999)\citenamefont
  {Benvenuti}, \citenamefont {Calatroni}, \citenamefont {Campisi},
  \citenamefont {Darriulat}, \citenamefont {Peck}, \citenamefont {Russo},\ and\
  \citenamefont {Valente}}]{benvenuti1999study}%
  \BibitemOpen
  \bibfield  {author} {\bibinfo {author} {\bibfnamefont {C.}~\bibnamefont
  {Benvenuti}}, \bibinfo {author} {\bibfnamefont {S.}~\bibnamefont
  {Calatroni}}, \bibinfo {author} {\bibfnamefont {I.}~\bibnamefont {Campisi}},
  \bibinfo {author} {\bibfnamefont {P.}~\bibnamefont {Darriulat}}, \bibinfo
  {author} {\bibfnamefont {M.}~\bibnamefont {Peck}}, \bibinfo {author}
  {\bibfnamefont {R.}~\bibnamefont {Russo}},\ and\ \bibinfo {author}
  {\bibfnamefont {A.-M.}\ \bibnamefont {Valente}},\ }\bibfield  {title}
  {\bibinfo {title} {Study of the surface resistance of superconducting niobium
  films at 1.5 {GH}z},\ }\href@noop {} {\bibfield  {journal} {\bibinfo
  {journal} {Physica C: Superconductivity}\ }\textbf {\bibinfo {volume}
  {316}},\ \bibinfo {pages} {153} (\bibinfo {year} {1999})}\BibitemShut
  {NoStop}%
\bibitem [{\citenamefont {Venturini~Delsolaro}\ \emph
  {et~al.}(2018)\citenamefont {Venturini~Delsolaro}, \citenamefont {Rosaz},\
  and\ \citenamefont {Sublet}}]{longmarch}%
  \BibitemOpen
  \bibfield  {author} {\bibinfo {author} {\bibfnamefont {W.}~\bibnamefont
  {Venturini~Delsolaro}}, \bibinfo {author} {\bibfnamefont {G.}~\bibnamefont
  {Rosaz}},\ and\ \bibinfo {author} {\bibfnamefont {A.}~\bibnamefont
  {Sublet}},\ }\bibfield  {title} {\bibinfo {title} {The long march of niobium
  on copper},\ }\href@noop {} {\bibfield  {journal} {\bibinfo  {journal}
  {{CERN} Courier}\ } (\bibinfo {year} {2018})}\BibitemShut {NoStop}%
\bibitem [{\citenamefont {Padamsee}(2020)}]{padamsee2020history}%
  \BibitemOpen
  \bibfield  {author} {\bibinfo {author} {\bibfnamefont {H.}~\bibnamefont
  {Padamsee}},\ }\bibfield  {title} {\bibinfo {title} {History of gradient
  advances in {SRF}},\ }\href@noop {} {\bibfield  {journal} {\bibinfo
  {journal} {arXiv preprint:2004.06720}\ } (\bibinfo {year}
  {2020})}\BibitemShut {NoStop}%
\bibitem [{\citenamefont {Ciovati}\ \emph {et~al.}(2005)\citenamefont
  {Ciovati}, \citenamefont {Flood}, \citenamefont {Grenoble}, \citenamefont
  {King}, \citenamefont {Kneisel}, \citenamefont {Morrone},\ and\ \citenamefont
  {Snyder}}]{ciovati2005temperature}%
  \BibitemOpen
  \bibfield  {author} {\bibinfo {author} {\bibfnamefont {G.}~\bibnamefont
  {Ciovati}}, \bibinfo {author} {\bibfnamefont {R.}~\bibnamefont {Flood}},
  \bibinfo {author} {\bibfnamefont {C.}~\bibnamefont {Grenoble}}, \bibinfo
  {author} {\bibfnamefont {L.}~\bibnamefont {King}}, \bibinfo {author}
  {\bibfnamefont {P.}~\bibnamefont {Kneisel}}, \bibinfo {author} {\bibfnamefont
  {M.}~\bibnamefont {Morrone}},\ and\ \bibinfo {author} {\bibfnamefont
  {M.}~\bibnamefont {Snyder}},\ }\href@noop {} {\emph {\bibinfo {title}
  {Temperature mapping system for single cell cavities}}},\ \bibinfo {type}
  {Tech. Rep.}\ (\bibinfo  {institution} {JLAB-TN-05-059},\ \bibinfo {year}
  {2005})\BibitemShut {NoStop}%
\bibitem [{\citenamefont {Piel}\ and\ \citenamefont
  {Romijn}(1981)}]{piel1981cern}%
  \BibitemOpen
  \bibfield  {author} {\bibinfo {author} {\bibfnamefont {H.}~\bibnamefont
  {Piel}}\ and\ \bibinfo {author} {\bibfnamefont {R.}~\bibnamefont {Romijn}},\
  }\bibfield  {title} {\bibinfo {title} {Temperature mapping on a
  superconducting {RF} cavity in subcooled {H}e},\ }\href@noop {} {\bibfield
  {journal} {\bibinfo  {journal} {Nucl. Instr. Meth}\ }\textbf {\bibinfo
  {volume} {190}},\ \bibinfo {pages} {257} (\bibinfo {year}
  {1981})}\BibitemShut {NoStop}%
\bibitem [{\citenamefont {Bernard}\ \emph {et~al.}(1981)\citenamefont
  {Bernard}, \citenamefont {Cavallari}, \citenamefont {Chiaveri}, \citenamefont
  {Haebel}, \citenamefont {Heinrichs}, \citenamefont {Lengeler}, \citenamefont
  {Picasso}, \citenamefont {Piciarelli}, \citenamefont {T{\"u}ckmantel},\ and\
  \citenamefont {Piel}}]{bernard1981experiments}%
  \BibitemOpen
  \bibfield  {author} {\bibinfo {author} {\bibfnamefont {P.}~\bibnamefont
  {Bernard}}, \bibinfo {author} {\bibfnamefont {G.}~\bibnamefont {Cavallari}},
  \bibinfo {author} {\bibfnamefont {E.}~\bibnamefont {Chiaveri}}, \bibinfo
  {author} {\bibfnamefont {E.}~\bibnamefont {Haebel}}, \bibinfo {author}
  {\bibfnamefont {H.}~\bibnamefont {Heinrichs}}, \bibinfo {author}
  {\bibfnamefont {H.}~\bibnamefont {Lengeler}}, \bibinfo {author}
  {\bibfnamefont {E.}~\bibnamefont {Picasso}}, \bibinfo {author} {\bibfnamefont
  {V.}~\bibnamefont {Piciarelli}}, \bibinfo {author} {\bibfnamefont
  {J.}~\bibnamefont {T{\"u}ckmantel}},\ and\ \bibinfo {author} {\bibfnamefont
  {H.}~\bibnamefont {Piel}},\ }\bibfield  {title} {\bibinfo {title}
  {Experiments with the {CERN} superconducting 500 {MH}z cavity},\ }\href@noop
  {} {\bibfield  {journal} {\bibinfo  {journal} {Nuclear Instruments and
  Methods in Physics Research}\ }\textbf {\bibinfo {volume} {190}},\ \bibinfo
  {pages} {257} (\bibinfo {year} {1981})}\BibitemShut {NoStop}%
\bibitem [{\citenamefont {Bernard}\ \emph {et~al.}(1980)\citenamefont
  {Bernard}, \citenamefont {Cavallari}, \citenamefont {Chiaveri}, \citenamefont
  {Haebel}, \citenamefont {Heinrichs}, \citenamefont {Lengeler}, \citenamefont
  {Picasso}, \citenamefont {Picciarelli},\ and\ \citenamefont
  {Piel}}]{bernard1980first}%
  \BibitemOpen
  \bibfield  {author} {\bibinfo {author} {\bibfnamefont {P.}~\bibnamefont
  {Bernard}}, \bibinfo {author} {\bibfnamefont {G.}~\bibnamefont {Cavallari}},
  \bibinfo {author} {\bibfnamefont {E.}~\bibnamefont {Chiaveri}}, \bibinfo
  {author} {\bibfnamefont {E.}~\bibnamefont {Haebel}}, \bibinfo {author}
  {\bibfnamefont {H.}~\bibnamefont {Heinrichs}}, \bibinfo {author}
  {\bibfnamefont {H.}~\bibnamefont {Lengeler}}, \bibinfo {author}
  {\bibfnamefont {E.}~\bibnamefont {Picasso}}, \bibinfo {author} {\bibfnamefont
  {V.}~\bibnamefont {Picciarelli}},\ and\ \bibinfo {author} {\bibfnamefont
  {H.}~\bibnamefont {Piel}},\ }\bibfield  {title} {\bibinfo {title} {First
  results on a superconducting {RF}-test cavity for {LEP}},\ }in\ \href@noop {}
  {\emph {\bibinfo {booktitle} {11th International Conference on High-Energy
  Accelerators}}}\ (\bibinfo {organization} {Springer},\ \bibinfo {year}
  {1980})\ pp.\ \bibinfo {pages} {878--885}\BibitemShut {NoStop}%
\bibitem [{\citenamefont {Pekeler}\ \emph {et~al.}(1996)\citenamefont
  {Pekeler}, \citenamefont {Proch}, \citenamefont {M{\"o}ller}, \citenamefont
  {Champion}, \citenamefont {Graber}, \citenamefont {Lipatov}, \citenamefont
  {Padamsee}, \citenamefont {Matheisen}, \citenamefont {Crawford},\ and\
  \citenamefont {Fuljahn}}]{pekeler1996thermometric}%
  \BibitemOpen
  \bibfield  {author} {\bibinfo {author} {\bibfnamefont {M.}~\bibnamefont
  {Pekeler}}, \bibinfo {author} {\bibfnamefont {D.}~\bibnamefont {Proch}},
  \bibinfo {author} {\bibfnamefont {W.}~\bibnamefont {M{\"o}ller}}, \bibinfo
  {author} {\bibfnamefont {M.}~\bibnamefont {Champion}}, \bibinfo {author}
  {\bibfnamefont {J.}~\bibnamefont {Graber}}, \bibinfo {author} {\bibfnamefont
  {L.}~\bibnamefont {Lipatov}}, \bibinfo {author} {\bibfnamefont
  {H.}~\bibnamefont {Padamsee}}, \bibinfo {author} {\bibfnamefont
  {A.}~\bibnamefont {Matheisen}}, \bibinfo {author} {\bibfnamefont
  {C.}~\bibnamefont {Crawford}},\ and\ \bibinfo {author} {\bibfnamefont
  {T.}~\bibnamefont {Fuljahn}},\ }\bibfield  {title} {\bibinfo {title}
  {Thermometric study of electron emission in a 1.3-{GH}z superconducting
  cavity},\ }\href@noop {} {\bibfield  {journal} {\bibinfo  {journal} {Part.
  Accel.}\ }\textbf {\bibinfo {volume} {53}},\ \bibinfo {pages} {35} (\bibinfo
  {year} {1996})}\BibitemShut {NoStop}%
\bibitem [{\citenamefont {Makita}\ \emph {et~al.}(2015)\citenamefont {Makita},
  \citenamefont {Ciovati},\ and\ \citenamefont
  {Dhakal}}]{makita2015temperature}%
  \BibitemOpen
  \bibfield  {author} {\bibinfo {author} {\bibfnamefont {J.}~\bibnamefont
  {Makita}}, \bibinfo {author} {\bibfnamefont {G.}~\bibnamefont {Ciovati}},\
  and\ \bibinfo {author} {\bibfnamefont {P.}~\bibnamefont {Dhakal}},\
  }\href@noop {} {\emph {\bibinfo {title} {Temperature mapping of
  nitrogen-doped niobium superconducting radiofrequency cavities}}},\ \bibinfo
  {type} {Tech. Rep.}\ (\bibinfo  {institution} {Thomas Jefferson National
  Accelerator Facility (TJNAF), Newport News, VA},\ \bibinfo {year}
  {2015})\BibitemShut {NoStop}%
\bibitem [{\citenamefont {Padamsee}(1983)}]{padamsee1983calculations}%
  \BibitemOpen
  \bibfield  {author} {\bibinfo {author} {\bibfnamefont {H.}~\bibnamefont
  {Padamsee}},\ }\bibfield  {title} {\bibinfo {title} {Calculations for
  breakdown induced by "large defects" in superconducting niobium cavities},\
  }\href@noop {} {\bibfield  {journal} {\bibinfo  {journal} {IEEE transactions
  on magnetics}\ }\textbf {\bibinfo {volume} {19}},\ \bibinfo {pages} {1322}
  (\bibinfo {year} {1983})}\BibitemShut {NoStop}%
\bibitem [{\citenamefont {Russenschuck}(2011)}]{russenschuck2011field}%
  \BibitemOpen
  \bibfield  {author} {\bibinfo {author} {\bibfnamefont {S.}~\bibnamefont
  {Russenschuck}},\ }\href@noop {} {\emph {\bibinfo {title} {Field computation
  for accelerator magnets: analytical and numerical methods for electromagnetic
  design and optimization}}}\ (\bibinfo  {publisher} {John Wiley \& Sons},\
  \bibinfo {year} {2011})\BibitemShut {NoStop}%
\bibitem [{\citenamefont {Conway}\ \emph {et~al.}(2017)\citenamefont {Conway},
  \citenamefont {Ge},\ and\ \citenamefont
  {Iwashita}}]{conway2017instrumentation}%
  \BibitemOpen
  \bibfield  {author} {\bibinfo {author} {\bibfnamefont {Z.}~\bibnamefont
  {Conway}}, \bibinfo {author} {\bibfnamefont {M.}~\bibnamefont {Ge}},\ and\
  \bibinfo {author} {\bibfnamefont {Y.}~\bibnamefont {Iwashita}},\ }\bibfield
  {title} {\bibinfo {title} {Instrumentation for localized superconducting
  cavity diagnostics},\ }\href@noop {} {\bibfield  {journal} {\bibinfo
  {journal} {Superconductor Science and Technology}\ }\textbf {\bibinfo
  {volume} {30}},\ \bibinfo {pages} {034002} (\bibinfo {year}
  {2017})}\BibitemShut {NoStop}%
\bibitem [{\citenamefont {Canabal}\ \emph {et~al.}(2007)\citenamefont
  {Canabal}, \citenamefont {Bowyer}, \citenamefont {Chacon}, \citenamefont
  {Gillespie}, \citenamefont {Madrid},\ and\ \citenamefont
  {Tajima}}]{canabal2007development}%
  \BibitemOpen
  \bibfield  {author} {\bibinfo {author} {\bibfnamefont {A.}~\bibnamefont
  {Canabal}}, \bibinfo {author} {\bibfnamefont {J.}~\bibnamefont {Bowyer}},
  \bibinfo {author} {\bibfnamefont {P.}~\bibnamefont {Chacon}}, \bibinfo
  {author} {\bibfnamefont {N.}~\bibnamefont {Gillespie}}, \bibinfo {author}
  {\bibfnamefont {M.}~\bibnamefont {Madrid}},\ and\ \bibinfo {author}
  {\bibfnamefont {T.}~\bibnamefont {Tajima}},\ }\bibfield  {title} {\bibinfo
  {title} {Development of a temperature mapping system for 1.3-{GH}z 9-cell
  {SRF} cavities},\ }in\ \href@noop {} {\emph {\bibinfo {booktitle} {2007 IEEE
  Particle Accelerator Conference (PAC)}}}\ (\bibinfo {organization} {IEEE},\
  \bibinfo {year} {2007})\ pp.\ \bibinfo {pages} {2406--2408}\BibitemShut
  {NoStop}%
\bibitem [{\citenamefont {Salerno}\ and\ \citenamefont
  {Kittel}(1997)}]{salerno1997thermal}%
  \BibitemOpen
  \bibfield  {author} {\bibinfo {author} {\bibfnamefont {L.~J.}\ \bibnamefont
  {Salerno}}\ and\ \bibinfo {author} {\bibfnamefont {P.}~\bibnamefont
  {Kittel}},\ }\href@noop {} {\emph {\bibinfo {title} {Thermal contact
  conductance}}},\ \bibinfo {type} {Tech. Rep.}\ (\bibinfo  {institution}
  {National Aeronautics and Space Administration, Ames Research Center},\
  \bibinfo {year} {1997})\BibitemShut {NoStop}%
\bibitem [{\citenamefont {Salerno}\ \emph {et~al.}(1984)\citenamefont
  {Salerno}, \citenamefont {Kittel},\ and\ \citenamefont
  {Spivak}}]{salerno1984thermal}%
  \BibitemOpen
  \bibfield  {author} {\bibinfo {author} {\bibfnamefont {L.}~\bibnamefont
  {Salerno}}, \bibinfo {author} {\bibfnamefont {P.}~\bibnamefont {Kittel}},\
  and\ \bibinfo {author} {\bibfnamefont {A.}~\bibnamefont {Spivak}},\
  }\bibfield  {title} {\bibinfo {title} {Thermal conductance of pressed copper
  contacts at liquid helium temperatures},\ }\href@noop {} {\bibfield
  {journal} {\bibinfo  {journal} {AIAA journal}\ }\textbf {\bibinfo {volume}
  {22}},\ \bibinfo {pages} {1810} (\bibinfo {year} {1984})}\BibitemShut
  {NoStop}%
\bibitem [{\citenamefont {Salerno}\ \emph {et~al.}(1994)\citenamefont
  {Salerno}, \citenamefont {Kittel},\ and\ \citenamefont
  {Spivak}}]{salerno1994thermal}%
  \BibitemOpen
  \bibfield  {author} {\bibinfo {author} {\bibfnamefont {L.~J.}\ \bibnamefont
  {Salerno}}, \bibinfo {author} {\bibfnamefont {P.}~\bibnamefont {Kittel}},\
  and\ \bibinfo {author} {\bibfnamefont {A.~L.}\ \bibnamefont {Spivak}},\
  }\bibfield  {title} {\bibinfo {title} {Thermal conductance of pressed
  metallic contacts augmented with indium foil or {A}piezon grease at liquid
  helium temperatures},\ }\href@noop {} {\bibfield  {journal} {\bibinfo
  {journal} {Cryogenics}\ }\textbf {\bibinfo {volume} {34}},\ \bibinfo {pages}
  {649} (\bibinfo {year} {1994})}\BibitemShut {NoStop}%
\bibitem [{\citenamefont {Piel}(1980)}]{piel1980diagnostic}%
  \BibitemOpen
  \bibfield  {author} {\bibinfo {author} {\bibfnamefont {H.}~\bibnamefont
  {Piel}},\ }\bibfield  {title} {\bibinfo {title} {Diagnostic methods of
  superconducting cavities and identification of phenomena},\ }in\ \href@noop
  {} {\emph {\bibinfo {booktitle} {Proc. 1st Workshop on {RF}
  Superconductivity, Karlsruhe, Germany}}}\ (\bibinfo {year} {1980})\
  p.~\bibinfo {pages} {85}\BibitemShut {NoStop}%
\bibitem [{\citenamefont {Smith}(1969)}]{smith1969review}%
  \BibitemOpen
  \bibfield  {author} {\bibinfo {author} {\bibfnamefont {R.}~\bibnamefont
  {Smith}},\ }\bibfield  {title} {\bibinfo {title} {Review of heat transfer to
  helium {I}},\ }\href@noop {} {\bibfield  {journal} {\bibinfo  {journal}
  {Cryogenics}\ }\textbf {\bibinfo {volume} {9}},\ \bibinfo {pages} {11}
  (\bibinfo {year} {1969})}\BibitemShut {NoStop}%
\bibitem [{\citenamefont {Bald}\ and\ \citenamefont
  {Wang}(1976)}]{bald1976nucleate}%
  \BibitemOpen
  \bibfield  {author} {\bibinfo {author} {\bibfnamefont {W.}~\bibnamefont
  {Bald}}\ and\ \bibinfo {author} {\bibfnamefont {T.-Y.}\ \bibnamefont
  {Wang}},\ }\bibfield  {title} {\bibinfo {title} {The nucleate pool boiling
  dilemma},\ }\href@noop {} {\bibfield  {journal} {\bibinfo  {journal}
  {Cryogenics}\ }\textbf {\bibinfo {volume} {16}},\ \bibinfo {pages} {314}
  (\bibinfo {year} {1976})}\BibitemShut {NoStop}%
\bibitem [{\citenamefont {Schmidt}(1981)}]{schmidt1981review}%
  \BibitemOpen
  \bibfield  {author} {\bibinfo {author} {\bibfnamefont {C.}~\bibnamefont
  {Schmidt}},\ }\bibfield  {title} {\bibinfo {title} {Review of steady state
  and transient heat transfer in pool boiling helium {I}},\ }in\ \href@noop {}
  {\emph {\bibinfo {booktitle} {Stability of superconductors in helium {I} and
  helium {II}}}}\ (\bibinfo {year} {1981})\BibitemShut {NoStop}%
\bibitem [{\citenamefont {Amrit}\ and\ \citenamefont
  {Francois}(2000)}]{amrit2000heat}%
  \BibitemOpen
  \bibfield  {author} {\bibinfo {author} {\bibfnamefont {J.}~\bibnamefont
  {Amrit}}\ and\ \bibinfo {author} {\bibfnamefont {M.}~\bibnamefont
  {Francois}},\ }\bibfield  {title} {\bibinfo {title} {Heat flow at the
  niobium-superfluid helium interface: Kapitza resistance and superconducting
  cavities},\ }\href@noop {} {\bibfield  {journal} {\bibinfo  {journal}
  {Journal of low temperature physics}\ }\textbf {\bibinfo {volume} {119}},\
  \bibinfo {pages} {27} (\bibinfo {year} {2000})}\BibitemShut {NoStop}%
\bibitem [{\citenamefont {Amrit}\ and\ \citenamefont
  {Ramiere}(2013)}]{amrit2013kapitza}%
  \BibitemOpen
  \bibfield  {author} {\bibinfo {author} {\bibfnamefont {J.}~\bibnamefont
  {Amrit}}\ and\ \bibinfo {author} {\bibfnamefont {A.}~\bibnamefont
  {Ramiere}},\ }\bibfield  {title} {\bibinfo {title} {Kapitza resistance
  between superfluid helium and solid: role of the boundary},\ }\href@noop {}
  {\bibfield  {journal} {\bibinfo  {journal} {Low Temperature Physics}\
  }\textbf {\bibinfo {volume} {39}},\ \bibinfo {pages} {752} (\bibinfo {year}
  {2013})}\BibitemShut {NoStop}%
\bibitem [{\citenamefont {Halbritter}(1970)}]{halbritter1970comparison}%
  \BibitemOpen
  \bibfield  {author} {\bibinfo {author} {\bibfnamefont {J.}~\bibnamefont
  {Halbritter}},\ }\bibfield  {title} {\bibinfo {title} {Comparison between
  measured and calculated {RF} losses in the superconducting state},\
  }\href@noop {} {\bibfield  {journal} {\bibinfo  {journal} {Zeitschrift
  f{\"u}r Physik}\ }\textbf {\bibinfo {volume} {238}},\ \bibinfo {pages} {466}
  (\bibinfo {year} {1970})}\BibitemShut {NoStop}%
\end{thebibliography}%

\end{document}